\documentclass[11pt,a4paper]{article}
\pdfoutput=1
\usepackage{jheppub}

\usepackage{amsmath}
\usepackage{verbatim}
\usepackage{graphicx}
\usepackage{mathrsfs}
\usepackage{appendix}

\newcommand{\e}{\epsilon}
\newcommand{\bL}{\bar{L}}
\renewcommand{\L}{{\mathcal{L}}}
\newcommand{\bcL}{\bar{{\mathcal{L}}}}
\newcommand{\z}{{\bar z}}
\newcommand{\h}{{\bar h}}

\newcommand{\be}[1]{ \begin{equation}\label{#1} }
\newcommand{\ee}{\end{equation}}
\newcommand{\bea}[1]{\begin{eqnarray}\label{#1} }
\newcommand{\eea}{\end{eqnarray}}
\newcommand{\bes}{\begin{subequations}}
\newcommand{\ees}{\end{subequations}}

\newcommand{\p}{\partial}

\newcommand{\refb}[1]{(\ref{#1})}

\renewcommand{\>}{\rangle}

\newcommand{\non}{\nonumber}
\newcommand{\lb}{\left[}
\newcommand{\rb}{\right]}

\title{Flat Holography: Aspects of the dual field theory}

\author[a, b]{Arjun Bagchi,} \author[c]{Rudranil Basu,} 
\author[d]{Ashish Kakkar,} \author[a, d]{and Aditya Mehra.} \author{\\}

\affiliation[a]{Indian Institute of Technology Kanpur, Kalyanpur, Kanpur 208016. INDIA. \\}

\affiliation[b]{Center for Theoretical Physics, Massachusetts Institute of Technology,\\ 77 Massachusetts Avenue, Cambridge, MA 02139, USA.\\} 

\affiliation[c]{Saha Institute of Nuclear Physics,  Block AF, Sector 1, Bidhannagar, Kolkata 700068. INDIA. \\}

\affiliation[d]{Indian Institute of Science Education and Research, Dr Homi Bhabha Road, Pashan. \\ Pune 411008. INDIA.\\} 

\emailAdd{abagchi@iitk.ac.in, rudranil@sinp.ac.in, ashish.kakkar6@gmail.com, adityams@iitk.ac.in}

\abstract{Assuming the existence of a field theory in $D$ dimensions dual to $(D+1)$-dimensional flat space, governed by the asymptotic symmetries of flat space, we make some preliminary remarks about the properties of this field theory. We review briefly some successes of the 3d bulk -- 2d boundary case and then focus on the 4d bulk -- 3d boundary example, where the symmetry in question is the infinite dimensional BMS$_4$ algebra. We look at the constraints imposed by this symmetry on a 3d field theory by constructing highest weight representations of this algebra. We construct two and three point functions of BMS primary fields and surprisingly find that symmetries constrain these correlators to be identical to those of a 2d relativistic conformal field theory. We then go one dimension higher and construct prototypical examples of 4d field theories which are putative duals of 5d Minkowski spacetimes. These field theories are ultra-relativistic limits of electrodynamics and Yang-Mills theories which exhibit invariance under the conformal Carroll group in $D=4$. We explore the different sectors within these Carrollian gauge theories and investigate the symmetries of the equations of motion to find that an infinite ultra-relativistic conformal structure arises in each case.}

\preprint{}

\begin{document}

\maketitle


\section{Introduction}

\subsection{The Strominger Triangle}
Of late, there has been a resurgence in the study of quantum gravity in asymptotically flat spacetimes. This renewed interest is primarily due to Strominger's new insights \cite{Strominger:2013jfa} into the apparently unrelated physics of asymptotic symmetries, soft theorems in quantum field theory and the so-called memory effects. A beautiful story has emerged linking these three corners with what we will call the ``Strominger triangle". 

In a theory of gravity, the symmetries at the boundary of a spacetime is given formally by the Asymptotic Symmetry Group (ASG). This is the group of all diffeomorphisms allowed by a particular choice of boundary condition modded out by the trivial diffeomorphisms. Often the ASG is just the isometry of the vacuum state, e.g. in AdS$_{D+1}$ for $D>3$, the ASG is $SO(2, D)$. But there are famous exceptions to this and perhaps one of the most remarkable is the analysis of Brown and Henneaux \cite{Brown:1986nw} who showed that in AdS$_3$, the asymptotic symmetry is enhanced to two copies of the infinite dimensional Virasoro algebra. This infinite enhancement of symmetries is the reason of the many miracles of the AdS$_3$/CFT$_2$ correspondence. 

In asymptotic flat spacetimes in four dimensions, there is again an infinite enhancement of asymptotic symmetries when one constructs the ASG at null infinity. The ASG is the infinite dimensional Bondi-Metzner-Sachs group \cite{ Bondi:1962px, Sachs:1962zza} and not the conventionally expected Poincare group. The Poincare group is extended by the so-called ``super-translations" which are translations along the null direction which depend on the angles of the sphere at infinity. The BMS group also has the Lorentz subgroup which acts conformally on the sphere at null infinity. Following arguments similar to \cite{Belavin:1984vu} in 2d conformal field theories, there is a further infinite enhancement called ``super-rotations" when we don't require the generators on the sphere to be well-defined \cite{Barnich:2010eb}. Since the BMS group acts on the null boundary of flat space, it is natural to expect that this arises as a symmetry in the gravitational S-matrix. This was surprisingly not investigated before the current spurt of activity. It was conjectured in \cite{Strominger:2013jfa} that this is indeed the case for an infinite-dimensional subgroup of the full BMS group, a certain combination of the group acting on $\mathscr{I}^+$ and the one acting on $\mathscr{I}^-$, which will be called the diagonal BMS group. 

Weinberg's soft graviton theorem \cite{Weinberg:1965nx} equates a S-matrix element of a quantum theory of gravity to another S-matrix element which differs by the addition of a graviton which is ``soft", i.e. whose four-momentum is taken to zero. This is a universal formula and surprisingly does not depend on the details of the other particles involved in the process. In \cite{He:2014laa}, it was shown that Weinberg's theorem could be understood as a Ward identity arising out of the supertranslational invariance of the conjectured symmetry under the diagonal BMS group. New soft-graviton theorems related to Ward identities arising out of diagonal super-rotation invariance of the gravitational S-matrix have also been discovered \cite{Cachazo:2014fwa}, leading ultimately to a Virasoro invariance of the quantum gravity S-matrix \cite{Kapec:2014opa}. 

The third vertex of the Strominger triangle is the memory effect \cite{Zeldovich}. The gravitation memory effect refers to the displacement of a pair of nearby detectors when a pulse of radiation passes through a certain region of spacetime where these detectors are placed. The pulse of radiation produces a gravitational field leading an oscillation in their relative positions. Once the waves have passed, the detectors in general do not get back to their initial positions and this leads to the above effect. The memory effect is connected to the soft theorems by Fourier transforms (in time) \cite{Strominger:2014pwa}. And the link between BMS symmetries and the memory effects is through vacuum transitions. The infinite dimensional BMS symmetry implies that the classical vacuum of the gravitational theory in asymptotically flat spacetimes is infinitely degenerate. Quantum mechanically this implies that the Minkowski vacuum is not invariant under supertranslations and hence the symmetry is spontaneously broken. The associated Goldstone bosons are the soft gravitons. The different vacua are related by super-translations and differ from each other by the addition of soft gravitons. The (standard) gravitational memory effect has been shown to be equivalent to a transition between an initial and final vacuum state \cite{Strominger:2014pwa}.  There also has been the discovery of a new gravitational memory effect, called the spin memory which are related to super-rotations \cite{Pasterski:2015tva}. These again are related to the new soft graviton theorems by a Fourier transform in the time direction. 

The above triangle of relations is not only true for gravity, but for any field theory and in particular, similar relations for electrodynamics have been discussed in \cite{He:2014cra}. We will not go into the details of these here. What we want to briefly comment about is perhaps one of the most interesting consequences of this emerging picture of infra-red physics in quantum field theories. In \cite{Hawking:2016msc}, Hawking, Perry and Strominger have proposed that these new insights into gravity in asymptotically flat spacetimes can have a bearing on the black hole information loss paradox.  
The famous no-hair theorems in general relativity state that stationary black holes are completely characterised by their mass, angular momentum and electric charge up to diffeomorphisms. The authors of \cite{Hawking:2016msc} argue however that since supertranslations are diffeomorphisms that change the physical state, they map the stationary black hole to a physically inequivalent one and hence all black holes carry an infinity of supertranslation charges. These charges may go a long way in a resolution of the information loss paradox. 

\subsection{Construction of holography for flatspace}

Having given a brief outline of the new excitement in the field of research of gravity in asymptotically flat spacetimes, let us now motivate our work in this paper. We are interested in a holographic formulation of flat space and this paper is a collection of results in this direction. 

The canonical asymptotic symmetry analysis in a theory of gravity leads to the idea of the ASG as we have discussed. We shall assume, following most known examples of holography, that this ASG also dictates the symmetries of a dual field theory that lives on the boundary of the spacetime. Hence, a holographic formulation of quantum gravity in asymptotically flat spacetimes would mean that the putative dual field theories are BMS invariant theories in one dimension lower than the gravity theory, living on the null boundary of Minkowski spacetime. This line of thought has been successful in dealing with the three dimensional theory of Minkowskian gravity. In this work, we want to report some preliminary progress in the formulation of the holographic dual theory in higher dimensional asymptotically flat spacetimes based on the idea above. We will thus be confining ourselves to the ASG vertex of the Strominger triangle discussed above. But there are tantalising prospects of attempting to understand what the other vertices mean for the putative dual field theory. We will not have any concrete suggestions about these in the present work.    

Flat spacetimes can be obtained as a limit from AdS when the radius of AdS is taken to infinity. This means that the asymptotic symmetry structure of AdS should also go over to the asymptotic structure of flat space in this singular limit. In an algebraic sense, this is a contraction and the above statement can be exemplified well in the case of three spacetime dimensions. For higher dimensions, the process of contraction does not give the full picture, but it still remains a useful tool for some quantities of interest.  This line of thought would remain a persistent one throughout the current work. We shall, in this work, provide some basic features like construction of the highest weight representation theory for BMS algebras and provide examples of BMS invariant field theories which would serve as prototypical examples of field theories dual to asymptotically Minkowskian spacetimes. 

\subsection{Outline of the paper}

Now let us briefly sketch the outline of the rest of the paper. We start in Sec 2 with comments on the formulation of holography for a generic spacetime and then particularly for asymptotically flat spacetimes. We then review the formulation for 3d flat holography and summarise some of the successes of the programme. This is also important to contrast the formulation in three dimensional bulk and the higher dimensions.  

In Sec 3, we focus on generalities of BMS$_4$ invariant field theories in 3 dimensions. These are theories which would be dual to 4d asymptotically Minkowski spacetimes and which live on the null boundary of flat space. We construct the highest weight representations of the infinite algebra and then compute the two and three point correlation functions of such theories. It surprisingly turns out that these correlation functions are exactly the ones obtained for relativistic 2d conformal field theories. 

We then move into higher dimensions. In Sec 4, after a look at the construction of the conformal Carroll algebra by an ultra-relativistic contraction of the relativistic conformal algebra in any dimensions, we propose a way to lift this finite dimensional algebra to an infinite one. For this we also discuss the similarities and differences of the CCA with another contraction of the conformal algebra, the Galilean Conformal Algebra and its corresponding infinite dimensional extension. We also discuss briefly the geometrical structures which are associated with these field theories. In Sec 5, we focus on the representation theory aspects of the CCA. We construct highest weight representations labelled by the dilatation and rotation generators. The role that the ultra-relativistic boosts play in this context is carefully examined.  

In Sec 6, we construct in detail the theory of ultra-relativistic Electrodynamics and ultra-relativistic Yang Mills theories. These theories, like their non-relativistic cousins Galilean Electrodynamics \cite{LBLL, Bagchi:2014ysa} and Galilean Yang-Mills \cite{Bagchi:2015qcw}, have various different sectors which depend on the way that the gauge fields are scaled. We look at the equations of motion and the contracted versions of gauge symmetry in these various limits. We then go on to show that in $D=4$, all these different sectors have Conformal Carrollian invariance and there is an emergent infinite dimensional symmetry, which is the one we had proposed for the CCA earlier in the paper. We conclude in Sec 7 with a summary of our results and discussions. There is an appendix which contains explicit details of the ultra-relativistic $SU(2)$ Yang-Mills theory.  

\section{Flat Holography: Generalities} 

To set the stage for our analysis of the field theory dual to flat space, we now make some general remarks about holography in general and in particular how to set up the same problem for asymptotically Minkowski spacetimes. We then recall some of the basic features of the construction in the case of the 3d bulk and 2d boundary theory, which so far is the best understood case. 

\subsection{The recipe for holography} 
One of the basic ingredients in the formulation of a holographic correspondence is the matching of symmetries between the bulk and boundary theories. A canonical analysis of the symmetries at the boundary of a space-time leads to the notion of the asymptotic symmetry group which, for a particular choice of boundary conditions, is the group of allowed diffeomorphism of a space-time modded out by the trivial ones. This asymptotic symmetry group then is also identified with the underlying symmetry of the field theory dual which lives on the boundary. 

Often, the asymptotic symmetry group is nothing but the isometry of the vacuum and this is the case for the well known examples of AdS$_{D+1}$ for $D>2$. The symmetry group in question then is $SO(D,2)$ and this also is the symmetry underlying the $D$ dimensional dual field theory. As is well known,  $SO(D,2)$ is the conformal group in $D$ dimensions and the dual field theory is a CFT. This very basic premise can be taken as the first building block for formulating a new holographic duality. This, e.g., has been the way dS/CFT was proposed in \cite{Strominger:2001pn} and more recently, the asymptotic analysis of Vasiliev theories in three dimensions, which results in two copies of $\mathcal{W}$ algebras \cite{Campoleoni:2010zq, Henneaux:2010xg, Gaberdiel:2010ar}, led to the formulation of the Gaberdiel-Gopakumar conjecture \cite{Gaberdiel:2010pz} for a duality for higher spin theories.

We want to understand how to construct the notion of holography for flat space times. To this end, let us construct the statements equivalent to the above for flat space. The asymptotic symmetries of Minkowski spacetimes at null infinity is given by the Bondi-Metzner-Sachs group. This is a group which is infinite dimensional in all dimensions, first discovered for $D=4$ by Bondi, van der Burg and Metzner \cite{Bondi:1962px}, and, independently, Sachs \cite{Sachs:1962zza} in the late 1960's. It was found that under suitable boundary conditions, the asymptotic symmetries at null infinity in four dimensional Minkowski spacetimes becomes infinite dimensional defying common wisdom that this analysis should yield the Poincare group. The null boundary of flat space is ${\rm I\!R} \times S^{D-2}$. The BMS group consists of conformal transformations of the sphere at infinity and infinite dimensional generalised translations of the null direction, called super-translations, which depend on the angular co-ordinates on this sphere. In dimensions $D\geq5$, one can consistently set boundary conditions to truncate the BMS group to the Poincare group \cite{Hollands:2003ie,Tanabe:2009va, Tanabe:2011es}, but this is not possible for $D=3,4$. 

Going by the conventional notion of holography, the dual field theory to Minkowski spacetimes should thus be one which lives on the null boundary and is endowed with the symmetries of the BMS group. For $D\geq5$, one can take the conservative point of view of the restricted boundary conditions to construct dual theories with Poincare symmetry. But this is much more exotic than it sounds, as a higher dimensional Poincare invariance needs to be imposed on a theory in one lower dimension. E.g. in the 5D bulk -- 4D boundary perspective, one would need to construct 4D field theories with symmetry $ISO(4,1)$. 

Adopting the conservative approach to flat holography mentioned above, in order to construct field theories can be candidates holographic duals to flat space, one can choose a systematic route. $ISO(D, 1)$ is a Inonu-Wigner contraction of $SO(D,2)$. So the $D$-dimensional field theories dual to $(D+1)$-dimensional flat space can be constructed as systematic limits of $D$-dimensional conformal field theories. It was shown in \cite{Bagchi:2012cy} that in terms of space-time directions, the contraction of the parent conformal field theory that one needs to consider is along the time direction of the theory. This physically amounts to sending the speed of light to zero in the field theory and hence can be viewed as an ultra-relativistic boost of the parent conformal field theory. 

With this in mind, later in the paper, we shall revisit one of the simplest systems to exhibit classical conformal invariance, viz. the Maxwellian theory of electromagnetism in $D=4$. Here we would find that an ultra-relativistic limit on the theory leads to a 5d Poincare invariant theory which lives in 4 dimensions. We will then propose a lift of the symmetry algebra to the infinite dimensional BMS$_5$ algebra and check the invariance of the theory under this infinite algebra. Our analysis would be very similar to the one carried out in \cite{Bagchi:2014ysa}. Then, following recent work in \cite{Bagchi:2015qcw}, we will also display the BMS$_5$ invariance of classical Yang Mills theory in $D=4$. 

It is important to mention that the above strategy of using the ultra-relativistic (UR) limit is one of the tools that we use in our analysis and not the only one. Another important one is the process of giving infinite dimensional lifts to the aforementioned finite dimensional algebra in the spirit of \cite{Bagchi:2009my}, where a similar infinite dimensional lift was carried out when exploring the non-relativistic (NR) limit of the conformal algebra. Checking how the infinite dimensional algebras form symmetries of the field theories we construct would be an important step that we will construct in our analysis. 

\subsection{Holography of 3d Minkowski spacetimes} 
Most of the recent work on holography in flat spacetimes has been carried out in three dimensions. If we look at the Asymptotic Symmetry Group at the null boundary of flat space, this now becomes the infinite dimensional BMS$_3$ group and the corresponding algebra is given by 
\begin{subequations}\label{GCA2}
\begin{align}
[ L_n, L_m ] &= (n-m) L_{m+n} + \frac{c_L}{12}  m(m^2-1) \delta_{m+n,0} \\
[L_n, M_m] &= (n-m) M_{m+n} + \frac{c_M}{12}  m(m^2-1) \delta_{m+n,0} \\
[M_n, M_m] &= 0
\end{align}
\end{subequations}
The null boundary $\mathscr{I}^\pm$ of 3d flat space has a structure of ${\rm I\!R} \times S^1$. In the above, $L_n$'s are the generators of the diffeomorphisms on the circle and the $M_n$'s are the so-called ``super-translations", translations along the null direction which depend on the angle. $c_L$ and $c_M$ are the possible central extensions of the algebra allowed by Jacobi identities. A canonical analysis finds them to be $c_L=0$ and $c_M = \frac{3}{G}$ in Einstein gravity \cite{Barnich:2006av}. 

It is well known that the ASG of AdS$_3$ is also infinite dimensional and following the seminal analysis of Brown and Henneaux \cite{Brown:1986nw} can be computed to be two copies of the Virasoro algebra 
\begin{subequations}
\begin{align}
[ \L_n, \L_m ] &= (n-m) \L_{m+n} + \frac{c}{12}  m(m^2-1) \delta_{m+n,0} \\
[ \bcL_n, \bcL_m ] &= (n-m) \bcL_{m+n} + \frac{\bar{c}}{12}  m(m^2-1) \delta_{m+n,0}, \quad [ \L_n, \bcL_m ] = 0.
\end{align}
\end{subequations}
Here $c=\bar{c}= \frac{3 \ell}{2G}$ where $\ell$ is the radius of AdS and $G$ is the Newton's constant. 
The BMS$_3$ algebra descends naturally from this structure when one takes the radius of AdS to infinity. This can be seen by looking at the following linear combinations
\be{}
L_n = \L_n - \bcL_{-n}, \quad M_n = \frac{1}{\ell} (\L_n + \bcL_{-n} ) 
\ee
It is easily seen that in this limit, one needs to combine the central terms as $c_L = c - \bar{c}$ and $c_M = \frac{1}{\ell} (c - \bar{c})$ and this reproduces the above expressions for the central terms given the AdS$_3$ Brown-Henneaux central terms.
 
Interestingly, the algebra \refb{GCA2} was also previously studied in the context of non-relativistic versions of CFTs \cite{Bagchi:2010zz} and these field theories went under the name of Galilean Conformal Field Theories (GCFTs) \cite{Bagchi:2009my}. The reader may be confused by the appearance of the term non-relativistic when we have just mentioned that the field theories we would be considering actually can be thought of as ultra-relativistic boosts of some parent CFT. The reason for this is the magic of two dimensions. Here a contraction in the space direction (a non-relativistic limit) is equivalent to a contraction in the time direction (an ultra-relativistic limit) in terms of the symmetry algebra. This magic does not hold in the higher dimensions where the number of directions in which the contraction would take place begin to differ. In a $(D+1)$ dimensional field theory, the non-relativistic limit would mean a contraction of the $D$ spatial directions as opposed to the ultra-relativistic limit which still remains a contraction in only one direction, time {\footnote{Note however that one can device skewed non-relativistic limits and hence have a more general notion of a BMS/GCA correspondence \cite{Bagchi:2010zz}.}}.  

The infinite dimensional symmetries of these field theories have allowed a number of applications in holographic studies of 3d flat space. We summarise some of the important ones briefly below. 
\begin{itemize}
\item{{\em{Cardy-like counting}} \cite{Bagchi:2012xr} (see also \cite{Barnich:2012xq}): } The infinite dimensional symmetries of the dual theory helps one employ techniques of usual CFTs to construct the partition function of these field theories and then with the notion of a modified modular invariance of these partition functions which emerges in the limit, one can write down a modified Cardy formula for these theories. 
The formula reproduces the Bekenstein-Hawking entropy of the so-called Flat space cosmologies (FSCs) which are (shifted-boost) orbifolds of flat-space and are time dependent solutions of Einstein's equations in asymptotically flat spacetimes with cosmological horizons \cite{Cornalba:2003kd}. The logarithmic corrections for the formula have been worked out in \cite{Bagchi:2013qva}. The modified Cardy formula has also been reproduced as a limit in \cite{Riegler:2014bia, Fareghbal:2014qga}. 

\item{{\em{Flat-space chiral gravity}} \cite{Bagchi:2012yk}: } This was the first explicit example of a bulk theory with a flat space boundary conditions and a known field theory dual. The bulk theory is a singular limit of Topologically Massive Gravity where the Gravitational Chern-Simons term becomes the only important term and the Einstein-Hilbert term drops off. It was shown that this was also the limit when the dual theory reduced to a single copy of the Virasoro algebra. The theory was called Flat-space chiral gravity (F$\chi$G). As an explicit example, it was conjectured that F$\chi$G at Chern-Simons level $k=1$ was dual to the $c=24$ Monster CFT.  A check for this conjecture is that the chiral Cardy formula reproduces the entropy of the FSC solutions in F$\chi$G \cite{Bagchi:2013hja}. 

\item{{\em{Flat limit of Liouville}} \cite{Barnich:2012rz}: }  An explicit example of a possible 2d field theory dual realising the whole of the BMS algebra was constructed in \cite{Barnich:2012rz}. The authors discussed two different flat limits of Liouville theory, one which described \refb{GCA2} without central extensions and one which had non-zero central terms.  

\item{{\em{Cosmic phase transitions}} \cite{Bagchi:2013lma}: } Phase transitions between FSCs and hot flat space was discovered in \cite{Bagchi:2013lma}. These are analogues to the Hawking-Page phase transitions in AdS with the interesting difference that these are transitions from time-independent uneventful flat space to a time-evolving cosmology. The formulation of a well-defined variational principle for 3d asymptotically flat space led to the discovery of an non-standard boundary term \cite{Detournay:2014fva} that was used in formulation of the phase transition discussed above.

\item{{\em{Higher spin theories}} \cite{Afshar:2013vka, Gonzalez:2013oaa}: } A higher spin version of the BMS story in three dimensions was discovered in \cite{Afshar:2013vka, Gonzalez:2013oaa}. The resulting asymptotic symmetry algebra (which we decided to call the BMW algebra!), initially obtained by the usual canonical analysis, can also be obtained as a contraction of analogous linear combinations of two copies of the corresponding $\mathcal{W}$ algebra.  Some follow-up work can be found in \cite{ Gary:2014ppa, Matulich:2014hea}. 

\item{{\em{Entanglement Entropy}} \cite{Bagchi:2014iea}: } Using techniques similar to those in 2d CFTs \cite{Calabrese:2004eu}, the entanglement entropy of 2d field theories with BMS symmetry (or Galilean CFTs) was computed in \cite{Bagchi:2014iea}. On the bulk side, with a computation using Wilson lines in the Chern-Simons formulation of gravity in three dimensions, the same answers were reproduced. More details about the computation were given in a follow-up paper \cite{Basu:2015evh}.  

\item{{\em{Stress Energy Correlations}} \cite{Bagchi:2015wna}: } Invoking the infinite dimensional algebra, it is possible to compute arbitrary n-point correlation functions of the energy-momentum tensor of these 2d Galilean conformal field theories by invoking the highest weight conditions on the representations. One can write down Ward identities corresponding to these infinite dimensional symmetries. Like in 2d CFTs, the connected part of these energy momentum n-point correlation functions follow recursion relations. The correlation functions and the recursion relations can be holographically derived by invoking the Chern-Simons formulation again and implementing the highest weight gauge in this formulation. 

\item{{\em{Induced representations}} \cite{ Barnich:2014kra, Barnich:2015uva, Campoleoni:2016vsh}: } The above checks of the holographic principle in flat space assumed that for holographic considerations, one needed to use highest weight representations of the underlying algebra. But the problem with these highest weight conditions is that the representations are generically non-unitary. Explicitly unitary representations of the BMS algebra have been considered and these are induced representations constructed in \cite{Barnich:2014kra, Barnich:2015uva}. They also have been recently constructed in the limit from AdS in \cite{Campoleoni:2016vsh}. It is likely that both highest weight and induced representations would have a role to play in the full construction of flat holography and each would be important for particular applications.  

\item{{\em{Non-Riemannian boundary structures}}: } The field theory dual to flat space lives on the null boundary of flat space. This means that the metric of the field theory is degenerate. The Riemannian nature of spacetime on which the field theory is defined is replaced and what replaces it is very similar to the Newton-Cartan structures found in non-relativistic spacetimes. The manifold is called a Carrollian manifold and in two dimensions, this is dual to the Newton-Cartan manifold \cite{Duval:2014uva}. These new structures are expected to play a central role in the formulation of Minkowskian holography. Some preliminary steps have been taken in \cite{Hartong:2015xda, Hartong:2015usd}. 

\end{itemize}

It is important to mention some short-comings of the above programme. As we have said, the appearance of the infinite dimensional BMS group is at the null boundary of flat space. In current investigations, one has only used one field theory living on either $\mathscr{I}^+$ or $\mathscr{I}^-$. For a complete description of holography in flat space times, one needs to consider a field theory living on the whole of the null boundary. This means we need to understand how to put together these two 2d GCFTs, one living on $\mathscr{I}^+$ and the other on $\mathscr{I}^-$. Along with that there is another problem. There exists a different asymptotic symmetry group, called the SPI group, at spatial infinity $i_0$. $i_0$ contains important information about gravitational charges like mass and angular momentum of the space-time and it is very likely that the field theory that lives at this point would also be important for flat holography. The effort so far of trying to use GCFTs to describe 3d flat space can be looked upon as only a partial attempt at solving what is clearly a problem with diverse complexities. Having these caveats in mind, we will now go on to address some of the basic features of flat holography in four bulk dimensions.

\newpage

\section{Holography of 4d Flat Space}
As we stressed in the introduction, the Poincare group in 4 dimensions comes from the contraction of the AdS$_4$ isometry group. So, if the Poincare group turned out to be the group of asymptotic symmetries of 4d Minkowski spacetimes, then the dual field theory would have the symmetries of the contraction of a 3d relativistic CFT. However, the group of asymptotic symmetries in 4d turns out to have a far richer structure and the dual theory, if there exists one, can only be partially understood from the point of view of a contraction of a relativistic CFT. In this section{\footnote{This section was worked out in collaboration with Anandita De}}, we first revisit the infinite symmetries of the putative dual theory of 4d flat space and then explore some basic aspects of its representation theory. We then go on to focus on only a particular Poincare sub-algebra which arises from the contraction of the AdS isometries to extract information of the two and three point functions of the 3D field theory in question.  

\subsection{Symmetries of BMS$_4$}
As we have mentioned in the beginning of the last section, Bondi, van der Burg and Metzner, and, independently, Sachs discovered in the late 1960's that under suitable boundary conditions, the asymptotic symmetries at null infinity in four dimensional Minkowski spacetimes becomes infinite and is given by the following algebra, which has then come to be known as the 4d Bondi-Metzner-Sachs (BMS$_4$) algebra:
\begin{subequations}\label{bms4}
\begin{align}
& [ L_n, L_m ] = (n-m) L_{m+n}, \quad [\bL_n, \bL_m] = (n-m) \bL_{n+m} \\
& [L_m, M_{r,s}] = \left( \frac{m+1}{2} - r \right) M_{m+r, s} \ , \quad [\bL_m, M_{r,s}] = \left( \frac{m+1}{2} - s\right) M_{r, m+s} \\
& [M_{r,s}, M_{t,u}] = 0
\end{align}
\end{subequations}
Here $n,m$ range from $-1$ to $+1$ while the other variables can take all integral values. The generators $M_{r,s}$ are called the super-translation generators which are translations that can depend on the angles of the sphere at infinity. The $L_n$ generators are the ones of the global conformal algebra of the sphere at infinity. In \cite{Barnich:2010eb}, it was proposed that like in 2d CFTs, if the restriction of being globally well defined is relaxed, the $L_n$ generators too could become infinite dimensional and the first line of the BMS algebra would then become two copies of the Witt algebra. We will work with this definition of the BMS$_4$, where both the $L$ and the $M$ generators are infinitely extended {\footnote{Here it is of interest to mention that the BMS$_4$ group has also been recently extended differently to include all smooth vector fields on $S^2$ instead of extending the global conformal symmetries to Virasoro symmetries \cite{Campiglia:2014yka, Campiglia:2015yka}. This generalisation of the BMS group is thus the semi-direct product of the supertranslations with Diff$(S^2)$, the group of smooth conformal deformations of the sphere at null infinity. We will not be working with this notion of the extended BMS group.}}.

We now come to the question of central extensions of \refb{bms4}. The literature seems to be a bit confused here. In \cite{Barnich:2011ct}, the authors concluded that the only non-trivial central terms that the algebra could have are the Virasoro central extensions to the Witt algebra of the first line. In \cite{Barnich:2011mi}, the same set of authors concluded that the Virasoro central terms were zero and there were non-zero, actually divergent, central term contributions in the cross commutators of $L$ and $M$. We performed a simple minded analysis of Jacobi identities of the generators of \refb{bms4} and agree with the initial result of possible central Virasoro extensions. Jacobi identities don't allow any central terms in the cross commutators{\footnote{Note that here unlike \cite{Barnich:2011mi} we are speaking of conventional central terms which are not field dependent.}}. 

\subsection{Highest weight representations}
We would like to construct the highest weight representation for the BMS$_4$ algebra in an attempt to understand the symmetry structure that underlies the possible 3d field theory dual to 4d Minkowski space times. The motivation behind doing this comes from the fact that this is the natural thing to do if one is to make connections to a limit from a parent CFT. Let us stress again that in the case of BMS$_4$ (which now would be the symmetry of a 3d field theory), the connection between the symmetry structures is only between the Poincare sub-group of the BMS$_4$ and the conformal algebra in 3d. The infinite dimensional nature of the BMS$_4$ algebra is something that cannot be easily understood from the limiting prescription. We shall have more to say about this later. 

Let us first choose a convenient representation for the BMS$_4$ algebra. We denote the directions in the field theory as $\{ x,y,t \}$ and use complex co-ordinates $z = x - iy$ and $\z = x+i y$. The generators are given by
\be{genbms4}
L_n = - z^{n+1} \p_z - \frac{1}{2} (n+1) z^n t \p_t, \quad \bL_n = - \z^{n+1} \p_\z - \frac{1}{2} (n+1) \z^n t \p_t, \quad M_{r,s} = z^r \z^s \p_t.
\ee
We will refer to this representation as the ``plane" representation. We choose to label the states of the 3d field theory under the $L_0, \bL_0$ generators:
\be{}
L_0 |\Phi\> = h |\Phi\>, \quad \bL_0 |\Phi \> = \h |\Phi\>
\ee
This particular choice of the labels is the physically relevant choice and this can be seen by considering the following linear combinations in the above representation:
\be{}
L_0 + \bL_0 = t \p_t + z \p_z + \z \p_\z= t\p_t + x \p_x + y\p_y , \quad L_0 - \bL_0 = z \p_z - \z \p_\z = i ( x \p_y - y \p_x)
\ee
So we see that these combinations are the dilatation operator and the rotation in the 3d field theory and $h+\bar{h}$ can be identified with the weight under the dilatation operator and $h -\bar{h}$ with the spin of the state under question. These are clearly the quantum numbers that one would like to associate with a state in the theory. The labelling of the states would be useful again when we connect later to the limit from 3d conformal field theories. 

Let us also point out another useful representation which we will call the ``cylinder" representation. The mapping between the ``cylinder" and ``plane" is given by:
\be{}
z = e^{i\xi}, \quad \z = e^{i \bar{\xi}}, \quad t = i \tau e^{i\xi} e^{i \bar{\xi}}
\ee
These names and indeed the two representations are motivated by their 2d counterparts discussed in \cite{Bagchi:2013qva}. In the cylinder co-ordinates, the generators take the following from:
\bea{}
&&L_n = i e^{i n \xi} \left[ \p_\xi + i \left(\frac{n-1}{2}\right) \tau \p_\tau \right], \quad  
\bL_n =i e^{i n \bar{\xi}} \left[ \p_{\bar{\xi}} + i \left(\frac{n-1}{2}\right) \tau \p_\tau \right], \\
&&M_{r,s} =-i e^{i (r-1) \xi} e^{i (s-1) \bar{\xi}} \p_\tau
\eea
The cylinder representations would be of use when we want to talk about BMS$_4$ invariant 3d field theories at finite temperature. 
It is important to note that since we have 
\be{}
[L_0, M_{0,0}] = \frac{1}{2} M_{0,0} = [\bL_0, M_{0,0}]
\ee
the states don't get any additional label under the $M_{0,0}$ generator. This is a very crucial point and in stark contrast to the highest weight representations of the BMS$_3$ (or GCA$_2$) algebra \refb{GCA2} where the states were defined by their labels under the single zero mode Virasoro generator ($L_0$) and also the zero mode super-translation generator ($M_0$) \cite{Bagchi:2009pe}. This aspect of the representation theory would make a very big difference between the physics of the field theories with BMS symmetry in 2 and 3 dimensions. We will have more to say about this when we are discussing some generic representation theory aspects in the next section.

We can build a notion of primary states in the BMS$_4$ algebra in close analogy with usual CFTs. Let us denote a state $\Phi$ with weights $(h, \h)$ as $ |h, \h\>$. We see that the generators $L_n, \bL_n, M_{n,m}$ for positive values of $m,n$ lower the weights of the state.   
\bea{}
&&L_{0}L_{n}|h , \h \rangle= (h-n)L_{n}|h, \h \rangle, \quad \bL_{0}\bL_{n}|h, \h\rangle =(\h-n)\bL_n |h, \h\>\\
&& L_{0}M_{r,s}|h, \h\rangle = (h- r+\frac{1}{2})M_{r,s}|h, \h\>,\quad
\bar{L}_{0}M_{r,s}|h, \h\rangle = (\bar{h} - s +\frac{1}{2})M_{r,s}|h, \h\rangle
\eea
So, we define primary states $ |h, \h\>_p$ as those which are annihilated by all generators $L_n, \bL_n, M_{n,m}$ with $m,n >0$. 
\be{}
L_n |h, \h\>_p = 0, \quad \bL_n |h, \h\>_p =0, \quad M_{r,s} |h, \h\>_p = 0 \qquad \forall \, n, r, s >0
\ee
The representation would be built by acting on these primary states by the creation operators $L_{-n}, \bL_{-n}, M_{-n,-m}$ with $m,n >0$. An interesting observation is that due to the commutation relations of $L$'s with $M$'s we now have half-integral as well as integral levels as opposed to the purely integral levels of the Virasoro algebra. 

It is now important to note that the 4d Poincare algebra is the 10 dimensional subgroup of the BMS algebra consisting of $\{L_{0, \pm1}, \bL_{0, \pm1}, M_{0,0}, M_{0,1}, M_{1,0}, M_{1,1}\}$. We would demand that the ground state of our 3d field theory which is the candidate dual to the 4d bulk Minkowski space-time is invariant under this sub-group of the BMS group and shall proceed to calculate correlation functions based on these symmetries. 


\subsection{2-pt and 3-pt functions}
We are now interested in calculating correlation functions of BMS$_4$ primaries constructed above. We shall use the symmetries of the Poincare sub-group of the BMS$_4$ as we mentioned. Using insight from CFTs, the proper thing to do is to use the globally well-defined part of the the BMS$_4$ algebra, which is the global part of the Witt algebra and the infinitely extended super translations. The reader may be concerned that the above restriction of using the Poincare sub-group instead of the other, much larger, group is contentious. We will dispel these doubts near the end of this section and show that the generators of super translations that we have ignored do not put any additional constraints on the correlators. 

We want the vacuum of the putative dual field theory to be invariant under the Poincare generators.
We wish to calculate the two point function of two BMS$_4$ primary fields $\Phi^i, \Phi^j$ which have weights $(h_1, \h_1)$ and $(h_2, \h_2)$. 
\be{}
G^{ij}(z_{1},\bar{z}_{1},t_{1}; z_{2}, \bar{z}_{2}, t_{2})=\langle 0| {\Phi^{i}(z_{1},\bar{z}_{1}, t_{1})\, \, \Phi^{j}(z_{2}, \bar{z}_{2},t_{2})}| 0 \rangle
\ee
Before looking at this, it is instructive to look at the form of the expressions of local operators at an arbitrary spacetime point. If $\Psi(z, \z,t)$ is a local operator, we have
\be{}
\Psi(\vec{x}) = U \Psi(0) U^{-1} \quad \mbox{with} \quad \vec{x} = (z, \z, t) \quad \mbox{and}\quad U = \exp{(t M_{0,0} - zL_{-1} - \z \bL_{-1})}
\ee
We would need local operators in our theory and the action of the generators of the algebra on them. For later use, we list a few expressions which can be obtained by the repeated use of the Baker-Campbell-Hausdroff formula.
\bea{BCH}
U^{-1}L_{n}U= \sum_{k=0}^{n+1} \frac{(n+1)! (-1)^{k} z^{k}}{k!}\frac{L_{n-k}}{(n+1-k)!}+\sum_{l=0}^{n} \frac{(n+1)! (-1)^{l}  z^{l}}{l!}\frac{t M_{n-l,0}}{2(n-l)!}\non
\\
U^{-1}\bar L_{n}U= \sum_{k=0}^{n+1} \frac{(n+1)! (-1)^{k} \bar{ z}^{k}}{k!}\frac{\bar L_{n-k}}{(n+1-k)!}+\sum_{l=0}^{n} \frac{(n+1)! (-1)^{l} \bar{z}^{l}}{l!}\frac{t M_{0,n-l}}{2(n-l)!}\non
\\
U^{-1}M_{nm}U= \sum_{l=0}^{m} \sum_{k=0}^{n} \frac{r!}{(r-l)!l!} \frac{s!}{(s-k)!k!} (-1)^{l+k} z^{l} \bar z^{k} M_{r-l,s-k}~~~~~~~~~~~~~~~~~~~~
\eea
Now let us return to the two point function of the two BMS primary operators. Using $L_{-1}, \bar{L}_{-1}, M_{00}$ we get that the two-point function is translationally invariant in all the three co-ordinates.
\be{}
G^{ij}(\vec{x_1}, \vec{x_2})= G^{ij}(z_{1}-z_{2}, \bar{z}_{1}-\bar{z}_{2}, t_{1}-t_{2}).
\ee 
Very interestingly, using $M_{01}, M_{10}$ we get $G^{ij}$ is independent of $t$. Using the above relations \refb{BCH}, when we determine the structure of the two-point function under the action of $L_0, \bL_0$, we get the following differential equations for the two point function
\begin{eqnarray}
(h_{1}+h_{2})G^{ij} + (z_1- z_2)\p_{z_1- z_2}G^{ij}=0 \\
(\h_1+\h_2)G^{ij} + ({\z_{1}}-{\z}_{2})\partial_{{\z_{1}}-{\z}_{2}}G^{ij}=0
\end{eqnarray}
Theses equations give us 
\begin{equation}
G^{ij} (\vec{x}_1, \vec{x}_2) =\dfrac{C^{ij}}{(z_{1}-z_{2})^{h_1 + h_2} (\z_1-\z_2)^{\h_1+\h_2}}
\end{equation}
The action of $M_{1,1}$ does not generate any extra constraints. From the action of $L_1$ and $\bL_1$, using the formulae \refb{BCH} we get
\begin{eqnarray}
(z_1^2 \partial_{z_1} +2 h_1z_1)G^{ij} + (z_2^2 \partial_{z_2} +2 h_2 z_{2})G^{ij}=0 \\
(\bar{z}_{1}^2\partial_{\bar{z}_{1}} +2{\h_{1}}\bar{z}_{1})G^{ij} + (\bar{z}_{2}^{2}\partial_{\bar{z}_{2}} +2\h_{2}\bar{z}_{2})G^{ij}=0
\end{eqnarray}
These equations give us $h_{1}=h_{2}$ and $\bar{h}_{1}=\bar{h}_{2}$. So we have
\begin{equation}
G^{ij} (\vec{x}_1, \vec{x}_2) = \dfrac{C^{ij}}{(z_1-z_2)^{2h} (\z_1 -\z_2)^{2\h}}
\end{equation}
This is identical to the two-point function of a {\em{two-dimensional relativistic conformal field theory}}. We shall see that the three point function also behaves in the same way. So we see that in the highest weight representation described above, there is a strong hint that the representation of the BMS$_4$ algebra reduce to that of two copies of the Virasoro algebra. This may lead us to speculate that the dual to 4d Minkowski spacetime may actually be a relativistic conformal field theory living in two dimensions.

As we said at the start of the section, one of the reasons for concern in this calculation may be that we have considered only the Poincare sub-group of the full BMS group and not the well-defined ``global" version which contains $L_{0, \pm1}, \bL_{0, \pm1}$ and all the super translations. We see that all the super translation generators have derivatives only in the time direction and crucially, the states are not labelled under the $M_{0,0}$ generator. Hence all the other generators do not impose any extra constraints on the two-point correlator. A pointer to this is the fact that we did not get any new constraint from the action of $M_{1,1}$ on the two-point function. The conclusion is that we would have ended up with the same two point function as we obtained above if we had considered the invariance under $L_{0, \pm1}, \bL_{0, \pm1}$ and all the super translations.

Let us now briefly mention the construction of the three point function of three BMS primary fields in our 3d field theory. The analysis is essentially identical to the one of the two-point function. The action of $L_{-1}, \bL_{-1}, M_{0,0}$ says that the three-point function depends on the differences of the co-ordinates and the action of $M_{1,0}, M_{0,1}$ gives us that the correlation function is independent of time. So we see again that the three-point function of a BMS$_4$ invariant field theory is just that of a conformal field theory in 2 dimensions. 

In the above, we have considered only scalar primary BMS operators. It would be of interest to check if the reduction also carries over to operators with spin. 
Another point of interest is that one can readily obtain these answers by looking at the ultra-relativistic contraction of the two and three point correlators in 3d relativistic conformal field theories. This is easily seen by looking at the expressions for the 2 and 3-point correlation functions of a usual 3d CFT and contracting the time co-ordinate. We will have more to say about the contraction of the time direction in the coming sections. This is intimately linked to the fact that the flat space limit of AdS is perceived as an ultrarelativistic limit on the dual field theory where, in a group theoretic sense, the time direction contracts.   

\paragraph{Supertranslations and vacuum degeneracy:} 
There is something very important that our analysis above assumes. It is the fact that the dual field theory to 4d asymptotically Minkowski spacetimes has a unique ground state. The vacuum of classical general relativity in 4d flat spacetimes is infinitely degenerate by the virtue of the infinite BMS group. Supertranslations generate transitions among these ground states and this is reflected in the memory effect as explained in the introduction. 

So how can we even begin to justify a dual theory which does not cover this basic feature? Our argument here is that even though the vacuum of classical general relativity is infinitely degenerate, the vacuum of quantum gravity in flat space is expected to be unique and Poincare invariant. This is the vacuum we are working with in the field theory dual. A unique CFT vacuum is also what we assume in the AdS/CFT correspondence. It is very likely that infrared issues similar to which arise in flat space, and that had been largely neglected in the literature in the past, would also arise when one includes a cosmological constant. Also, we would expect that the flat limit of AdS/CFT should lead to flat holography and hence the unique AdS vacuum would also go over to a unique vacuum for quantum gravity in flat space. Having argued our point, it would however be interesting in trying to formulate a way to understand the classical vacuum to vacuum transition and hence the memory effect in terms of the dual theory. We don't have anything concrete to say about this in the present work.    

\bigskip 

\bigskip

In conclusion, in this section we have shown that the two and three point functions of a 3d field theory invariant under the BMS$_4$ group reduce to those of a relativistic conformal field theory in one lower dimension. So there is the hint that the dual theory to 4d flat spacetime is a 2d relativistic conformal field theory. This rather simple analysis, based on the highest weight representations of the BMS$_4$ algebra, is reminiscent of some recent work \cite{Kapec:2014opa, Kapec:2016jld, Cheung:2016iub} where bulk physics of 4d asymptotically Minkowski spacetime was recovered from 2d CFTs. We hope to understand the link between the two approaches in the near future.

It is possible that when restricting attention to highest weight representations of the BMS$_4$ algebra, there is a consistent truncation to the modules of two copies of the Virasoro algebra. This is similar to the case in the BMS$_3$ algebra \refb{GCA2} where if one looks at the case where the central term $c_M$ and the weight of a primary state under the action of $M_0$, $h_M$,  are both zero.  In the context of the gravitational theory, this is what has been called Flat Space Chiral Gravity and we discussed this briefly in the review part of Sec 2. In the context of BMS$_4$, this would provide further evidence to the claim that the field theory dual to 4d asymptotic flat spacetimes is actually a 2d CFT. It would be instructive here to do a thorough analysis of the null vectors of the full BMS$_4$ algebra and check whether there is a reduction of two copies of the Virasoro algebra building on the lines of the same analysis in the BMS$_3$ case carried out in \cite{Bagchi:2009pe}. Our preliminary analysis here ran into computational hurdles due to the complexity of the underlying symmetry algebra. We hope to report on this issue in upcoming work. 

\newpage

\section{Conformal Carroll Groups: Contractions and infinite extensions.}

In this section, we begin explorations into higher dimensions. We begin by recalling the ultra-relativistic contraction of the conformal algebra in any dimension. This gives us the realisation of the conformal version of the Carroll algebra, the Conformal Carroll Algebra (CCA), in all dimensions. It is interesting and useful to contrast this with the non-relativistic contraction of the conformal algebra which results in the Galilean Conformal Algebra (GCA). The GCA can be given an infinite dimensional lift in all dimensions. We wish to explore similar infinite lifts for the CCA. The conformal Carrollian group has been shown to be isomorphic to the BMS groups in \cite{Duval:2014uva}. Building on this, we proceed to show how to get the infinite dimensional algebras presented in the previous sections, viz. BMS$_3$ and BMS$_4$, from the finite version of the CCA . We then outline a procedure to give an infinite lift in all dimensions, explicitly demonstrating the case of BMS$_5$. 

\subsection{Ultra-relativistic contractions} 
We shall construct a ultra relativistic limit of the conformal algebra by performing an In{\"o}n{\"u}-Wigner contraction. Here, we would take the limit at the level of the spacetime and the contraction will be performed in units where the speed of light $c=1$ and achieved if we do the following systematic operation:
\be{urc} 
x^{i} \rightarrow x^{i},~~ t \rightarrow \varepsilon t, \quad \varepsilon \to 0. 
\ee
The generators of ultra relativistic limit can be constructed out of the relativitsic conformal algebra by performing the contraction (\ref{urc}): 
\begin{eqnarray} \label{ura}
&& J_{ij}= -(x_{i}\partial_{j}-x_{j}\partial_{i}),\hspace{.2cm} D=-(t\partial_{t}+x^{i}\partial_{i}),\hspace{.2cm}K= (x^{k}x_{k})\partial_{t} \non \\ 
&& B_{i}= x_{i}\partial_{t}, \hspace{.2cm}H=\partial_{t},\hspace{.2cm} P_{i}=\partial_{i},\hspace{.2cm}
K_{i}=- 2x_{i}(t\partial_{t}+x^{k}\partial_{k})+(x^{k}x_{k})\partial_{i}
\end{eqnarray}
The non-trivial brackets among the generators are given by:
\bea{cca-finite} 
&& [J_{ij},B_{k}]=\delta_{k[i}B_{j]}, \hspace{.2cm}[J_{ij},P_{k}]= \delta_{k[i}P_{j]},\hspace{.2cm}   
[B_{i},P_{j}]=-\delta_{ij}H,\hspace{.2cm} [D,K]=-K, \non\\ &&\hspace*{.2cm} [K,P_{i}]=-2B_{i}, \hspace*{.2cm}[K_{i},P_{j}]=-2D\delta_{ij}-2J_{ij}, \hspace*{.2cm} [H,K_{i}]=-2B_{i}\\
&&[D,H]=H, \hspace*{.2cm}[D,P_{i}]=P_{i}, \hspace*{.2cm} [D,K_{i}]=-K_{i} \non.
\eea

It is instructive to note here that $\mathcal{G} = \{J_{ij}, D, P_i, K_i \}$ form one lower dimensional relativistic conformal algebra. The relation to the BMS algebra is very clear from here. The structure at null infinity is ${\rm I\!R} \times S^{d-2}$.  $\mathcal{G}$ is the conformal group on the sphere at infinity. The other generators $\mathcal{H}= \{H, B_i, K \}$ are the non-compact generators which give rise to the translational part of Poincare group. We will now proceed to give these an infinite dimensional extension to $\mathcal{H}$ to enhance this to the abelian supertranslation group.  But before that we would like to draw inspiration from a similar construction for the non-relativistic limit performed in \cite{Bagchi:2009my}. 

\subsection{Comparing with Galilean Conformal Algebra}
It is interesting to compare and contrast the above symmetry structure of the UR sector with the Galilean one \cite{Bagchi:2009my}, which captures the physics of the non-relativistic regime. From the point of contractions, it is clear that here one scales the spatial and temporal coordinates in a way opposite to \eqref{urc}, i.e.:
\bea{gal_lim}
x^i \rightarrow \varepsilon x^i, ~~ t \rightarrow t, \quad \varepsilon \to 0.
\eea
In this limit, we can organise the generators of the contracted relativistic conformal algebra, called the Galilean Conformal Algebra \cite{Bagchi:2009my}, into the following the set of vector fields:
\bea{suggen}
&& L^{(n)} = -t^{n+1} \p_t - (n+1) t^n x_i \p_i, \quad ( L^{(-1, 0, +1)} = H, D, K) \non\\
&& M^{(n)}_i = t^{n+1} \p_i, \quad ( M^{(-1, 0, +1)}_i = P_i, B_i, K_i) \\
&& J_{ij} =  x_i \p_j - x_j \p_i \non
\eea
for $n=0, \pm 1$. The interesting brackets are
\be{GCA}
[L^{(n)}, L^{(m)}] = (n-m) L^{(n+m)}, \quad [L^{(n)}, M^{(m)}_i] = (n-m) M_i^{(n+m)}, \quad  [M^{(n)}_i, M^{(m)}_j] =0 
\ee
and $J$ acts usually as a rotation generator. \eqref{GCA} is a another Inonu-Wigner contraction of the conformal algebra. Possibly the most interesting feature of this algebra is that, as the form suggests, the $n=0, \pm 1$ set is a sub-algebra of an infinite dimensional one, for which $n \in \mathbb{Z}$ and the latter is also named Galilean conformal algebra{\footnote{It is of interest here to note that the GCA admits a further central extension where the rotation generators are also lifted, viz. 
\be{}
J_{ij}^{(n)} = t^n (x_i \p_j - x_j \p_i)
\ee
The algebra then takes the form of an affine Kac-Moody algebra augmented by the algebra which we have above \refb{GCA} \cite{Bagchi:2009my}.}}. This infinite dimensional lift is obvious if we are in $D=2$ as the algebra there arises as a contraction of two copies of the Virasoro algebra as described before in Sec 2. But in higher dimensions, the appearance of an infinite symmetry structure is very counter-intuitive. At first this seems to be just some mathematical jugglery. But it can be shown that these infinite dimensional symmetries arise in the non-relativistic limit of physical systems, e.g. in non-relativistic limits of Electrodynamics \cite{Bagchi:2014ysa} and Yang-Mills theories \cite{Bagchi:2015qcw} in $D=4$.

\subsection{Infinite extension of the Conformal Carroll Algebra}
We look at how to extend the finite dimensional algebra \refb{cca-finite} to get infinite extensions to the BMS algebras in various dimensions inspired by the methods outlined above. First we will remind the reader what the extension to obtain BMS$_3$ and BMS$_4$ are, following \cite{Bagchi:2012cy}, and then will propose an extension for all dimension. 

Starting out from our intuition from the infinite lift of the GCA in arbitrary dimensions, we encounter an apparent road-block when we now look at the finite CCA. Generators of GCA are expressed as polynomials in Galilean time $t$. But, $t$ and $x^i$ switch roles as far as Galilean and ultra-relativistic scaling are concerned. Now, we want to inquire what tensorial structure the new generators with polynomials in $x^i$ should have. We already have scalar ($D,H$ etc.) and vector ones  ($B_i, P_i$ etc.) with highest power of $x$ being 2. Problems arise while introducing higher powers of $x^i$ in the generators through terms like $S_j = x^i x_i x_j \p_t$. When we attempt to construct the extended algebra with these new generators, the commutators of the new generators with the older ones no more close among themselves. The commutators in question give rise to newer generators with higher powers in $x$ and/or assign higher rank tensor structure than rank 1. In particular, the algebra thus constructed never closes at any finite polynomial of $x^i$. 

We start off trying to rectify this apparent hurdle by first addressing the two known cases. In $D=2$ (see for example \cite{Bagchi:2009pe}), the problem is simple, as $x$ does not carry a vectorial index. Here we can give the algebra an infinite lift by just exchanging $t\leftrightarrow x$ in \refb{suggen}: 
\be{}
L_n = -x^{n+1} \p_x - (n+1) x^n t \p_t, \quad M_n = x^{n+1} \p_t.
\ee
Here it is understood that the finite generators are the ones with $n=0, \pm 1$ and we extend the vector fields for all values on $n$ after this rearrangement. This leads to the same algebra as the 2d GCA (\refb{GCA} without the vectorial index on the $M_n$ generators). This isomorphism is down to the fact that the algebra does not distinguish between a contraction in the spatial direction and one in the temporal direction. 

For the $D=3$ case, the exercise is a bit more involved \cite{Bagchi:2012cy}. The trick is to correctly combine the finite generators, use complex co-ordinates and then look at the infinite extension. We make the following identifications (we have two spatial co-ordinates $x, y$ and time $t$ in the 3D field theory):
\bea{fin-genbms4}
&& L_0 = \frac{1}{2}(D+ i J_{xy}), \quad L_{-1} = -\frac{1}{2}(P_x + iP_y), \quad L_1 = \frac{1}{2}(K_x - i K_y); \non\\
&& \bar{L}_0 = \frac{1}{2}(D - i J_{xy}),\quad \bar{L}_{-1} = -\frac{1}{2}(P_x - iP_y), \quad \bar{L}_1 = \frac{1}{2}(K_x + i K_y); \\
&& M_{00} = P_0, \quad M_{01}= J_{0x} + i J_{0y}, \quad M_{10}= J_{0x} - i J_{0y}, \quad M_{11}=K_0 \non.
\eea
The above vector fields generate the algebra \refb{bms4}. The generalised form of the vector fields that generate this algebra is given by \refb{genbms4}. This coincides with \refb{fin-genbms4} for $n, m= 0, \pm1$ and $r, s = 0, 1$. If we now use the form of \refb{genbms4} for all integral values of the labels, we see that we generate the same algebra, now for all $\rm Z$. This is the infinitely extended BMS$_4$ algebra. 

\bigskip

The recurring feature in the two above example is that the BMS algebra contains a sub-algebra which is the conformal structure of the sphere at null infinity. The infinite dimensional structures that arose as the conformal structure in the BMS$_3$ was Diff$(S^1)$ which is one copy of the Virasoro and in the case of BMS$_4$, the conformal structure was extended to include all the generators of two copies of the Virasoro algebra, like the case of a 2d CFT. For $D>4$, the sphere at infinity is $S^n$ for $n>2$. The conformal structure is thus finite dimensional. We don't expect any infinite enhancement of the symmetry algebra from this part of the algebra. We have also mentioned that one can choose boundary conditions such that the ASG for flat spacetimes at $D>4$ reduces to the Poincare algebra. Why then are we interested in attempting to extend the contracted algebra for higher dimensions? 

This boils down to the recent investigations of the relations between the Weinberg soft theorem and the BMS group for higher dimensions. Weinberg's theorem is not dependent on dimensions and holds for $D>4$. This means that the theorem would lead to an infinite number of Ward identities which will naturally lead to the the expectation of an infinite dimensional supertranslation sub-group. This has been recently discussed in \cite{Kapec:2015vwa}{\footnote{The sub-leading soft theorems lead to the Ward identities for the superrotations. These have not been investigated for $D>4$. If such theorems are found, we would possibly need to consider other extensions of the BMS group like the one mentioned earlier in footnote 2.}}. With this in mind, we explore the possible infinite lift of 
the BMS group in higher dimensions. 

\bigskip

We will specialise now to 4d theories (3 spatial + 1 time direction) with the understanding that our construction can be generalised to arbitrary dimensions. In 3 spatial dimensions, arbitrary rank tensors constructed out of polynomials in $x^i$ can be arranged as irreducible representations of $SO(3)$. This can be best understood by the Lie brackets of those generators with $J_{ij}$. This hints at the possibility of introducing infinite lifts of the time translation generators though arbitrary and all possible tensorial polynomials of arbitrary spins. These can be written in terms of countable modes:
\be{}
M^{n_1, n_2, n_3} = x^{n_1} y^{n_2} z^{n_3} \p_t,
\ee
with $n_i $ taking values in all integers. Notice that this is in contrast to the GCA where the whole of the non-relativistically contracted conformal algebra gets embedded into the infinite dimensional modes $L^{(n)}, M^{(m)}_i$. In this ultra-relativistic case, only an Abelian ideal consisting of $H, B_i$ and $K$ fits into the infinite extension:  
\bea{}
&& H= M^{0,0,0}, \quad B_x = M^{1,0,0}, \\ 
&& K = M^{2,0,0} + M_{0,2,0}+ M^{0,0,2}.
\eea 
Their Lie brackets with the components of other generators are:
\bea{urc2}
  \left[P_{x},M^{m_{1},m_{2},m_{3}}\right] &=& m_{1}M^{m_{1}-1,m_{2},m_{3}}  \non \\
 \left[D,M^{m_{1},m_{2},m_{3}}\right] &=& -(m_{1}+m_{2}+m_{3}-1)M^{m_{1},m_{2},m_{3}} \non \\
  \lb K_{x},M^{m_{1},m_{2},m_{3}}\rb &=& -(m_{1}+2m_{2}+2m_{3}-2) M^{m_{1}+1,m_{2},m_{3}} \\
 && + m_{1}(M^{m_{1}-1,m_{2}+2,m_{3}}+M^{m_{1}-1,m_{2},m_{3}+2})   \non \\
 \lb J_{xy},M^{m_{1},m_{2},m_{3}}\rb &=& -m_{2} M^{m_{1}+1,m_{2}-1,m_{3}}+ m_{1} M^{m_{1}-1,m_{2}+1,m_{3}}  \non
\eea
Lie brackets with the other components of $P_i, K_i$ and $J_{ij}$ are obvious. We must admit that for aesthetic reasons, we could have recombined linearly the $M$ generators so that they transform as irreducible representations of the rotation group generated by $J$. However, that is not necessary for our present purpose.

\subsection{Carrollian viewpoint of UR symmetries}
There is another approach to look at the above conformal symmetry structure pertaining to UR systems. This is rather intrinsic to these systems and does not rely on the corresponding structures of Minkowski space-time, from which we took the above limits. Once we pronounce our idea of the UR regime through \eqref{urc}, its easy to see that the familiar structure of Riemannian geometry is not the correct geometric setting to work with. If one does not want to bring in gravitational effects and prefers staying in flat background, this translates into giving up Minkowski space. Because, the scaling necessarily makes the Minkowski metric degenerate. This feature is not special to Minkowski, but it is ubiquitous in all Riemann manifolds, if we zoom in to the UR sector. The natural framework for dealing with these situations where the metric tensor degenerates along some specific direction is Carroll manifolds \cite{Duval:2014uoa, Duval:2014uva, Duval:2014lpa}.

In general one equips a differential manifold with a couple of structures, namely a symmetric rank-2 covariant `metric' tensor $g$, which is positive semi-definite everywhere and a vector field $\xi$, which is degenerate direction of the metric \footnote{For our present discussion, we can do away with a possible connection structure which induces parallel transport.}. This is the definition of a Carroll manifold.

The next step would be to see whether we can have analogues of conformal isometries on a Carroll manifold. The most obvious way to define these conformal Carroll isometries is as diffeomorphisms which preserve the Carrollian structure, ie $g$ and $\xi$ upto conformal factors.

For exemplification let us consider the $d$ dimensional flat-Carroll space-time where in a suitable Cartesian coordinate system:
\bea{no}
 g = \delta_{ij} dx^i \otimes dx^j  ~~~ \mbox{and} ~~~ \xi = \p_t \quad \mbox{where } i,j =1, \cdots , d-1 \non
\eea
In this coordinate system, diffeomorphisms satisfying conformal Carroll isometries can be nicely put together as a vector field \cite{Duval:2014uva}
\bea{carr_diff} 
X = p^i P_i + \omega^{ij} J_{ij} + \delta D + k_i K^i + f(x^i)H 
\eea
where $ P_i, J_{ij}, D, K_i \mbox{ and } H $ are same as those appearing in \eqref{ura} These were earlier found by taking appropriate scaling and limiting prescription from the Minkowski ones. The rest of the quantities appearing in \eqref{carr_diff} are coefficients of linear combination, all of which are constant, except for the function $f$. This functional dependence makes the set of allowed diffeomorphisms an infinite dimensional group. For some specific choices of $f$ we observe that the generators $f (x^i) \p_t$ can be easily identified with the other (an Abelian ideal part) generators showed in \eqref{ura}):
\bea{carroll_gen2}  f = 1 \Rightarrow H,  ~~~ f = x^i \Rightarrow B_i   , ~~~ f = x^i x_i \Rightarrow K.   \non \eea
The other infinite number of modes $M^{m_{1},m_{2},m_{3}}$ are supposed to exhaust the set of all possible generators of the form $f(x^i) \p_t$. We note here that while all smooth vector fields $f \p_t$ are allowed for this purpose, we extended the our finite algebra \eqref{urc} through the modes $M^{m_{1},m_{2},m_{3}}$ which are obviously not globally smooth for negative $m_i$'s. Otherwise the two pictures are equivalent.

From now on, we will call the Lie algebra of the generators $P_i, D, K_i, J_{ij}$ together with all of the $M^{m_{1},m_{2},m_{3}}$'s, the infinite conformal Carroll algebra (i-CCA). 

\section{Highest weight representations of Infinite CCA}
In this section, we investigate some representation theory aspects of the i-CCA focussing on the highest weight representations. The most obvious ones to construct are the representations which are labelled by the dilatation operator and the spin. We shall also see why we cannot construct another representation, labelled by the dilatation and the boost generators and this will be another point of difference with the GCA in arbitrary dimensions.   
 
\subsection{The scale-spin representation}
We will be interested in the highest weight representation of the i-CCA, in a fashion similar to that for the case of 2D CFT. With this in mind, we label our states with definite scaling dimension $\Delta$ and spin $j$ (under $SU(2)$, as we are working for 3 spatial dimensions here ):
\be{wt}
D | \Phi \rangle = \Delta | \Phi \rangle, \quad
J^2 | \Phi \rangle = j(j+1) | \Phi \rangle
\ee
where we have $J^2 = J^2_{xy}+J^2_{yz} + J^2_{zx} $ as usual. The motivation behind choosing these as choice of highest weights stems from our intended investigation of the symmetries of gauge theories at the UR sector. In the ordinary relativistic setting, many gauge theories enjoy conformal invariance at least at the classical level. That's precisely why in our forthcoming analysis we would want dynamical fields to have definite scaling dimension. Moreover the UR limit takes Lorentz invariance $SO(D-1,1) \rightarrow ISO(D-1)$, while in Carrollian viewpoint $ISO(D-1)$ is manifestly a symmetry group. The spatial $SO(D-1)$ subgroup is always a symmetry of the theory and this helps us organise fields in the theory according to their spins. Much of our present analysis is going to be in direct analogy with the analysis that we carried out in \cite{Bagchi:2014ysa, Bagchi:2015qcw}. 

In what follows, we will interchangeably be talking in terms of states and fields. This means prescribing a state-operator correspondence analogous to 2D CFT:
\bea{}
\lim _{x^i, t \rightarrow 0 } \Phi (x^i,t) |\mathrm{vacuum} \rangle  =  | \Phi \rangle
\eea
For example, the equations \eqref{wt} therefore can alternatively be written as:
\be{altwt} 
\lb D, \Phi (0,0) \rb = \Delta \Phi (0,0),  \quad \lb J^2 , \Phi (0,0) \rb = j(j+1)  \Phi (0,0) 
\ee
Note that these brackets would mean commutators from now on as the generators have been promoted to quantum operators. At this point we would not explore the possibility of central extensions. Hence the map of the Lie algebra from the classical level of Lie brackets (e.g. \eqref{urc}) to commutators of quantum operators is an isomorphism. 

We construct the notion of a Carrollian primary state in analogy to that of a CFT. In view of the above commutators, the following criteria are natural for a primary operator to satisfy:
\be{condn}
\lb K_i, \Phi(0,0)\rb = 0, \quad \lb M^{m_{1},m_{2},m_{3}}, \Phi(0,0) \rb = 0 ~~~ \mbox{for at least one of }m_i > 1.
\ee
In addition, any field, including primary fields, at a generic point in spacetime transforms under the Hamiltonian and momentum operators as the following
\be{}
\lb H, \Phi (x^i,t) \rb = \p_t \Phi (x^i,t),  \quad \lb P_i, \Phi (x^i,t) \rb = -\p_i \Phi (x^i,t) 
\ee
Note that we have not specified the behaviour of the primary field under the action of the Carroll boosts $B_x = M^{1,0,0}$ etc. However this cannot be completely arbitrary. Hence it should be interesting to see how this action gets constrained by the consistency of CCA.

Keeping in mind our intended application to UR gauge theories descending from relativistic theories of vector gauge fields we keep ourselves limited to a multiplet of a $SU(2)$ scalar and a vector, say $\phi$ and $\phi_i$ and we would denote by these the fields evaluated at the origin. Consistency, in terms of Jacobi identities involving these fields with $B_i$ and $J_{ij}$,  leaves us with the following possibilities:
\be{}
\lb B_i, \phi \rb = a \, \phi_i, \quad \lb B_i, \phi_j \rb = b \, \delta_{ij} \phi
\ee
for some constants $a,b$, yet to be determined. As $B_i$ commutes with the dilatation operator, $\phi$ and $ \phi_i$ share same conformal dimension $\Delta$. A given $\Delta$ will therefore define a CCA primary multiplet.

For the complete information about the transformation properties of these primaries under CCA generators we would need explicit evaluation of commutator brackets at finite space-time points away from origin. The space and time translation operators come handy in defining these:
\bea{}
\Phi (x^i,t) = U \Phi (0,0) U^{-1} ~~ \mbox{where }~~ U = \exp (tH - x^i P_i).\non
\eea
The action of a symmetry generator on a CCA primary can therefore be seen as:
\bea{non_origin}
\delta _{\mathcal{O}} \Phi (x^i, t) &=& [\mathcal{O}, \Phi(x^i t)] = U [U^{-1} \mathcal{O} U, \Phi (0,0)] U^{-1}.
\eea
Using the algebra \eqref{urc} and \eqref{urc2}, we simplify $U^{-1} \mathcal{O} U$ and arrive at the desired results: 
\bes \label{symm_action}
\bea{}
&&\label{boost-sc-vect} [B_{i},\phi (x^i,t)] = x_{i}\partial_{t}\phi+ a\phi_{i},\quad [B_{i},\phi_{j} (x^i,t)] = x_{i}\partial_{t}\phi_{j} + b\, \delta_{ij}\phi \\
&& [K_{i},\phi (x^i,t)]= (2\Delta x_{i}+ 2x_{i}t\partial_{t}+2x_{i}x^{j}\partial_{j} -x^{j}x_{j}\partial_{i})\phi +2a\, t\phi_{i} \\
&& [K_{i},\phi_{l}(x^i,t)]= 2(\Delta x_{i} +x_{i}t\partial_{t}+x_{i}x^{j}\partial_{j}- \frac{x^{j}x_{j}}{2} \partial_{i})\phi_{l} +2x^{j}\delta_{li}\phi_{j} -2x_{l}\phi_{i} + 2b \, t \delta_{il}\phi \non\\
&& [ D, \phi (x^i,t)] =  (t \partial _t +x^i \partial _i + \Delta) \phi, \quad [ D, \phi _j(x^i,t)] =  (t \partial _t +x^i \partial _i + \Delta) \phi_j	\\
&& \label{M_on_scalar} [M^{m_{1},m_{2},m_{3}},\phi(x^i,t)]= x^{m_{1}}y^{m_{2}}z^{m_{3}}\p_{t}\phi +a\ \phi_i \ \p_i (x^{m_{1}}y^{m_{2}}z^{m_{3}})\\
&& \label{M_on_vector}[M^{m_{1},m_{2},m_{3}},\phi_{i}(x^i,t)]= x^{m_{1}}y^{m_{2}}z^{m_{3}}\p_{t}\phi_{i} + b\, \p_i (x^{m_{1}}y^{m_{2}}z^{m_{3}})\phi.
\eea
\ees
Derivations of \eqref{M_on_scalar} and \eqref{M_on_vector} warrant some discussion. These actions rely on the evaluation of $U^{-1} M^{m_{1},m_{2},m_{3}} U$. Employing, as usual, the Baker-Campbell-Hausdroff formula for expanding the exponentials in $U$, one arrives, for non-negative $m_1, m_2, m_3$, at the following finite sum:
\bea{}
U^{-1}M^{m_{1},m_{2},m_{3}}U =
\sum^{m_{1},m_{2},m_{3}}_{k,l,n=0} x^{k}y^{l}z^{n} \frac{m_{1}!m_{2}!m_{3}!~M^{(m_{1}-k)(m_{2}-l)(m_{3}-n)}}{k!l!n!~(m_{1}-k)!(m_{2}-l)!(m_{3}-n)!} 
\eea
While this implies validity of the formulas \eqref{M_on_scalar} and \eqref{M_on_vector} for non-negative modes, we would take this opportunity to extend it for all integers.

Additionally one should take notice of the fact that our representation of CCA is defined via the quadruplet $\{\Delta, j, a, b \}$. We will see in Sec.~6, focusing on particular field theories, how the representation characterizes a particular field describing a sector of UR physics.

\subsection{An alternative representation: scale-boost }
It is understandable that while describing physics where conformal invariance or scale transformation is an important aspect, assigning fields definite scaling dimension is crucial. Now, the structure of the conformal Carroll algebra reveals that the Carroll boost generators $B_i$ commute with scaling generator $D$. Hence an alternative to the scale-spin representation detailed above is a representation with fields having definite boost as well as scaling dimension. Similar representations were constructed for the Galilean conformal algebra \cite{Bagchi:2009ca} for arbitrary spatial dimensions. One of the interesting features of the scale-boost representations of the GCA was the existence of correlation functions with non-trivial dependence on both space and time directions. It is thus instructive to investigate whether similar interesting physical quantities can be extracted from the analogous representation of CCA. To make notations clear let's define a new primary state $| \Phi \rangle^{\prime}$ such that:
\be{}
D | \Phi \rangle^{\prime} = \Delta | \Phi \rangle^{\prime}, \quad B_i | \Phi \rangle^{\prime} = \xi_i | \Phi \rangle^{\prime}
\ee
for a constant vector $\xi_i$. We thus consider the notations $|\Phi \rangle^{\prime}$ and $| \Delta, \xi \rangle$ equivalent. One then sets up natural criteria analogous to \eqref{condn} for this state to be of highest weight. Proceeding along same lines, while evaluating the action of operators like $K_i$ on primary fields away from origin as done in \eqref{non_origin}, we encounter that:
\be{KJ}
U^{-1} K_i U = K_i + 2t\, B_i + x_i \,D + 2 x^j J_{ij}+2\, t x_i H - 2 x_i x^j P_j + x^j x_j P_i  
\ee
where, as usual, $U = \exp (tH - x^i P_i)$. From here we immediately identify that a knowledge of the action of the rotation generators $J_{ij}$ on the primaries is necessary. This is in sharp contrast to the case of GCA and this traced back to the difference between the two algebras. The important brackets that are of concern here is that of special conformal generators $K_i$ and momenta. In GCA, they commute, whereas in the Carrollian or the ultra-relativistic avatar, the result is:
$$ [K_i, P_j ] = -2 D \delta _{ij} - 2 J_{ij}.$$
Presence of the rotation generators here shows up above \refb{KJ}. Note also that this term would be absent when we are looking at 2 dimensions where there can be no notion of spatial rotation. So we can use the scale-boost representations for the 2d CCA and this goes back to the discussion of the isomorphism between the 2d GCA and 2d CCA. 

A similar situation arose in the scale-spin representation discussed above, where an information about the action of the boost operator on the primaries were necessary. We sorted that out by prescribing a generic action of $B_i$ on a given primary through:
$$\delta_{B_i} \Phi = \bigoplus_j \phi_{(j)},$$
and spreading through all irreducible representations labelled by $j$ of $SU(2)$. Afterwards, imposing physical motivations relevant to the problem at hand we restricted the above to finite sums. This readily indicates that we should describe the required action of the rotation generator using the spectrum of the $B_i$ operator. That being unknown, we write the action as a formal sum:
\bea{suggested}
J^2 |\Delta , \xi \rangle = \sum_{\xi^{\prime}} C^{(\xi^{\prime})} |\Delta, \xi^{\prime}\rangle 
\eea
over the spectrum of $B_i$. These states can also include descendants of the corresponding primary. Acting from left by $B$ twice on this, with some algebraic manipulations give:
\bea{}
8 \xi \cdot \xi |\Delta, \xi \rangle = \sum_{\xi'} (\xi' - \xi)\cdot (\xi' - \xi) C^{(\xi^{\prime})} |\Delta, \xi^{\prime}\rangle 
\eea
Here $\xi \cdot \xi$ etc. stand for the inner product defined via 3 dimensional trivial metric $\delta_{ij}$.

Assuming that $B$ should have orthogonal eigenstates now ensures that $\xi = 0$ and $C^{(\xi ')}_{ij} =0 \, \forall \xi ' \neq 0$. This implies, for \eqref{suggested}:
\bea{}
J^2 | \Delta, 0 \rangle = C^{(0)} | \Delta, 0 \rangle
\eea
which is again a eigen-equation for the spin operator. It follows that we don't have any non-trivial representation labelled by the weights scale and boost. In view of the above, we will be working only with the previously developed representation with definite scale and spin.

\newpage

\section{Ultra-relativistic regime of Gauge theories}
In this section, we come to the main discussion of this part of the paper, $i.e.$ conformal Carrollian invariance of Electrodynamics and Yang Mills (YM) theories in the ultra-relativistic limit. Electrodynamics without matter coupling is a free theory and studying its symmetries sets the stage for more involved gauge theories. More precisely, this analysis will self consistently help us determine the representations ($\{\Delta, j, a ,b \}$) corresponding to specific Carrollian fields. For non-Abelian examples, in view of avoiding a cumbersome analysis while keeping things to minimal non-triviality, we are going to consider the pure gauge theory without matter coupling or supersymmetry. 

Since conformal invariance in the ultra-relativistic set-up is one of our key objectives in the present study, our motivation behind looking at Electrodynamics and pure YM theory stems from the fact that these are classically relativistically conformally invariant in 4 dimensional space-time. It is thus expected that the finite conformal Carrollian invariance would emerge in the UR limit of these theories. We will go on to show that there is further infinite dimensional enhancement of these symmetries to the i-CCA (or the corresponding BMS group). YM theories describe dynamics of Minkowski space vector fields. Accordingly, we need to concentrate on Carrollian descendants of Minkowskian vector fields.

\subsection{Carrollian gauge fields}
As in the Galilean case, going to the UR limit means first breaking Lorentz invariance by splitting and treating the spatial and temporal components of a four vector differently under the limiting prescription. While the space and time coordinates are scaled ( $x^i \rightarrow x^i, ~~ t \rightarrow \epsilon \, t \mbox{ with } \epsilon \rightarrow 0$) the components of vector fields should scale appropriately. Note that while transiting to the Galilean framework \cite{LBLL} from the relativistic one, 4-vectors are taken to two distinct limits depending upon the causal nature of the parent Minkowskian vector. For example the tuple of a scalar and the 3-vector originating from a time-like Minkowski vector is named as Electric type while the one coming from a space-like one \footnote{Obviously this nomenclature is based on a Galilean formulation of Electrodynamics. We will, for historic consistency, continue using these names.} falls in the Magnetic category.

Hence the effective starting point of UR gauge theories should be the following limits on a Minkowski 4-vector $V_{\mu}$:
\bes \label{lims}
\bea{}
\mbox{Electric limit:} \quad V_t \to V_t, \, V_i \to \e V_i \label{elimcont}\\
\mbox{Magnetic limit:} \quad V_t \to \e V_t, \, V_i \to V_i \label{mlimcont}
\eea
\ees
One can now readily write the transformation properties of both these sectors under UR boost, the crucial transformation that indicates departure from relativistic physics. This would amount to taking the boost components of the relativistic Lorentz transformation on 4-vectors (ie. $\delta_{\mathrm{Lorentz}} V_{\mu} = x_{[\rho} \p _{\nu]} V_{\mu}+\eta_{\mu [\rho} V_{\nu]}$) and taking the $\e \rightarrow 0$ limit after the appropriate scaling \eqref{lims}:
\bes \label{lim_boost}
\bea{}
\mbox{Electric limit :} \quad &&\delta_{\mathrm{Boost}} V_t = x_i \p_t V_t, \quad \delta_{\mathrm{Boost}} V_i = x_j \p_t V_i + \delta_{ij} V_t. \\
\mbox{Magnetic limit :} \quad &&\delta_{\mathrm{Boost}} V_t = x_i \p_t V_t + V_i, \quad \delta_{\mathrm{Boost}} V_i = x_j \p_t V_i.  
\eea
\ees
Comparing these with the transformation properties derived purely from the representation of CCA \eqref{boost-sc-vect}, \eqref{M_on_vector}, we see that electric limit corresponds to $a =0, b=1$ whereas the magnetic one is for $a=1, b=0$.
\subsection{Symmetries of ultra-relativistic Electrodynamics}
One of the defining features of of a Carroll manifold is the existence of a degenerate metric. This makes it hard to construct an action principle for Carrollian invariant theories. However this issue takes us to a completely different set of enquiries, less directly connected to the present goal of our paper{\footnote{This problem of constructing actions for theories on manifold with degenerate metrics arises also in the Galilean case. It is important here to mention that there have been some attempts to construct actions for Galilean Electrodynamics $e.g.$ \cite{Bergshoeff:2015sic}.}. In an attempt to study the dynamical features of self-interacting gauge theories in the ultra-relativistic sectors, we therefore start with the relativistic Maxwell equations of motion and construct those in the UR regime, consistently taking limits. Hence, for each of the sectors (electric and magnetic) we will land up to one Carrollian scalar and one vector one. We will scale the fields as:
\bea{edcontra1}
A_t \rightarrow  A_t, \, && \, A_i \rightarrow  \e A_i
\eea for the electric sector and (cf. \eqref{mlimcont}):
\bea{edcontra2}
A_t \rightarrow \e A_t, \, && \, A_i \rightarrow  A_i
\eea
for the magnetic one. The resulting equations of motion are:
\bes \label{edeom}
\bea{} 
\label{urseleom} && \p^{i}\p_{i}A_{t}-\p^{i}\p_{t}A_{i}=0,~~ \p_{t}\p_{i}A_{t}-\p_{t}\p_{t}A_{i}=0; \\
\label{ursmageom} && \p^{i}\p_{t}A_{i}=0,~~ \p_{t}\p_{t}A_{i}=0.
\eea
\ees
One of the major goals of this work is to see whether gauge theories in the UR limit respects the Carrollian conformal symmetry. Our approach towards checking invariance of an equation of motion of the form:
\bea{}
f(A, \p A, \p^2 A) = 0
\eea
with respect to a particular symmetry generator $ \mathcal{O}$ would be to see whether the variational derivative equation
$$ \delta_{\mathcal{O}} f(A, \p A, \p^2 A) =0$$
holds. The variational actions are given as $\delta_{ \mathcal{O}}A = [\mathcal{O}, A]$ and the explicit expressions are given in \eqref{symm_action}. While checking invariance with respect to the space-time translations and spatial rotations are straightforward, those for the other CCA transformations are interesting. 

This is because, as we have mentioned earlier, a particular field on our flat Carroll manifold is specified by a quadruplet $\{\Delta, j, a ,b \}$. Only for spatial dimensions $D=3$, these come out to be invariant under dilatation transformation, as the case for the ordinary relativistic theory. Moreover let us remind the reader once more that in the Carrollian framework, we will mean by scalars the fields transforming as $j=0$ representation of the spatial rotation group and vectors transforming as $j=1$, (cf.\eqref{wt}). This on the other hand implies $\Delta = \frac{D-2}{2} = 1$ for $A_0$ and $A_i$ alike. The other two parameters $a,b$ that specify a particular representation also dictates the `electric' or `magnetic' nature of it. Values of these parameters were derived by looking at UR boost actions found from the appropriate Lorentz ones by limits. 
Once we have fixed these values, we can now check for the invariance of the equations of motion under the full infinite dimensional CCA. Let us elucidate this with an example: we check for the transformation of one component of the vector equation of \eqref{urseleom} under the infinite Carrollian supertranslations $M^{m_1, m_2, m_3}$:
\bea{} 
&& \delta_{M_{m_{1},m_{2},m_{3}}}\left(\p_{t}\p_{x}A_{t}-\p_{t}\p_{t}A_{x}\right)= (1-b)m_{1}x^{m_{1}-1}y^{m_{2}}z^{m_{3}}\p_{t}\p_{t}A_{t} + \\
 && a\,m_{1}\p_{x}(x^{m_{1}-1}y^{m_{2}}z^{m_{3}}\p_{t}A_{x})+ a\, m_{2}\p_{x}(x^{m_{1}}y^{m_{2}-1}z^{m_{3}}\p_{t}A_{y})+ a\, m_{3}\p_{x}(x^{m_{1}}y^{m_{2}}z^{m_{3}-1}\p_{t}A_{z}). \non
\eea
We see that is invariant if and only if $a=0, b=1$, precisely the set that defines Electric limit \eqref{lim_boost}. Similar invariance can be shown for $D$ and $K_i$ in these representations for the equation in the magnetic limit as well. This result can be summarily described by saying that Electrodynamics in the ultra-relativistic limit, in both the electric and magnetic sectors, has infinite conformal Carroll symmetry.

\bigskip

\subsection{Dynamics and symmetries of Yang Mills theory in UR regime}
Yang Mills theories are of course much richer in structure than Electrodynamics, as they are self-interacting non-linear theories. The dynamical content of ordinary relativistic non-Abelian YM theory, including the obvious gauge redundancy is contained in the Lie algebra valued 4-vector gauge potentials $A_{\mu}= A^a_{\mu} T_a$. Here $T_a \, (a=1, \dots , \mathfrak{D})$ are the Lie algebra generators corresponding to the $\mathfrak{D}$ dimensional gauge group. With this new structure of Lie algebra, there are more than just two ultra-relativistic limits of a YM theory. Hence we would like to understand the various possible of UR limits before we start working on the dynamics of gauge theories.

\subsubsection{Construction of different UR sectors}

Depending upon the projections onto electric like or magnetic like UR limits of the Lie algebra components, we would have in total $ \mathfrak{D} +1$ sectors, each being designated by one of these $\mathfrak{D}$ vectors:
\bea{}
\Xi_{(p)} = (\underbrace{0,0,\dots , 0}_{\mathfrak{D}-p}, \underbrace{1,1,\dots , 1}_{p}) \qquad ~~ p=0, \dots, \mathfrak{D} 
\eea
Additionally, $\Xi_{(p)}^a$ will denote its $a$'th component. This notation equips us with a more systematic handle while using the limits \eqref{lims} in YM equations of motion. Let's concentrate on a particular sector, say $p_0^{\mathrm{th}}$ sector, defined by the vector $\Xi_{(p_0)}$. The limits on the Lie algebra components the gauge field are conveniently captured by these definitions:
\bea{}
A^a_t \longrightarrow \frac{\epsilon}{1+\epsilon-\Xi^a_{(p_0)}}A^a_t,		~~~~ 	A^a_i \longrightarrow \frac{\epsilon}{\epsilon+\Xi^a_{(p_0)}}A^a_i.
\eea
Clearly if $\Xi_{(p_0)}^a=1$, the projection to the (Carroll) scalar and vector parts of the original gauge field components is (cf. \eqref{elimcont}):
\bea{ymcontra1}
A^a_t \rightarrow  A^a_t, \, && \, A^a_i \rightarrow  \e A^a_i
\eea
and we will call this electric-like scaling keeping in mind the nomenclature attached to electrodynamics \eqref{edcontra1}. 
On the other hand we get the magnetic-like analogue \eqref{edcontra2} here as well for $\Xi_{(p_0)}^a=0$ (cf. \eqref{mlimcont}):
\bea{ymcontra2}
A^a_t \rightarrow \e A^a_t, \, && \, A^a_i \rightarrow  A^a_i
\eea
We have just observed that different Lie algebra components are projected onto different UR limits (in the Carrollian language transform differently under CCA). Hence it would be nice to have Lie algebra components falling under electric and magnetic limits denoted respectively as $A^{\alpha}_t, A^{\alpha}_i$ and $A^{I}_t, A^{I}_i$. This means, the indices $a,b$ in the range $1, \dots, \mathfrak{D} - p_0$ will now be denoted by capital Romans $ I,J$ etc. and for the rest, ie, $\mathfrak{D} - p_0+1 \leq a \leq \mathfrak{D}$ Greekss $\alpha, \beta$ etc. will be used.

The familiar source-less equations of motion are:
\bea{rel_YM}
\p^{\mu} F_{\mu \nu}^a + g\, f^{a}{}_{bc} A^{\mu b} F^c_{\mu \nu}=0  \quad \mbox{where}  \quad F_{\mu \nu}^a = \p_{[\mu} A^a_{\nu]} + g \, f^a{}_{bc} A^b_{\mu} A^c_{\nu}. 
\eea
Here $g$ is Yang Mills self coupling, which is dimension-less in 4 dimensional space-time and $f$ are the Lie algebra structure constants. Now, depending upon the range of $a$ in equation \eqref{rel_YM}, there are two separate class of equations in its UR projection written below. 
\subsection*{Case 1: $1 \leq a \leq \mathfrak{D}-p_0$}
Let's look at the Carrollian scalar equation first:
\bea{remnant1}
\p^i \p_t A_i^I + g f^{I}{}_{JK} A^{iJ} \p_{t}A_i^{K}=0.
\eea
On the other hand, the vector one is:
\bea{remnant2}
\p _t \p_t A_j^I =0
\eea
\subsection*{Case 2: $\mathfrak{D}-p_0+1 \leq a \leq \mathfrak{D}$}
Scalar equation:
\bea{special}
f^{\alpha}{}_{IJ}A^{iI}\p_{t} A_{i}^{J}=0
\eea
Vector equation:
\bea{special2}
\p_t( \p_t A_j^{\alpha} - \p_j A_t^{\alpha}+g f^{ \alpha}{}_{\beta I} A_t^{\beta} A_j^{I}) +  g f ^{\alpha}{}_{ \beta I} A_{t}^{\beta}\p_{t} A^I_j=0
\eea

Few observations and caveats regarding these equations of motion are called for. We will discuss them in a coherent order.
 Firstly we note that equation \eqref{special} lacks kinetic term. Only remnant part after the UR projection is an interaction term. Apparently it would be wise to consider those sectors only, for which the equations of motion are \eqref{remnant1} and \eqref{remnant2} and accordingly set $p_0 =0$. And setting $p_0=0$ leaves one only with the magnetic sectors. It is crucial to observe that for these cases, there is no electric-magnetic leg mixing in the equations, unlike the equations \eqref{special} and \eqref{special2} where electric type fields like $A^\alpha$ mix as well with the magnetic-type ones. This seems to be in contrast to our expectation from the Electrodynamics case, where two distinct limits of electric and magnetic type survive after UR projection.

For this, let us take a detour along the procedure of constructing UR equations of motion from their relativistic avatars. The left hand side of a relativistic equation of motion 
$$F(A, \p A, \p^2 A) =0 $$ after appropriately scaling fields and coordinates yield equations of the form
\bea{expansion}
F(A, \p A, \p^2 A) \rightarrow \sum _{n=-k_1}^{k_2} f^{(n)} (A, \p A, \p^2 A) \e^n 
\eea
where $k_1 \leq k_2$ are integers. Natural regularization involves multiplying the last expression by $\e^{k_1}$ and take the limit $\e \rightarrow 0$. This is necessary to extract the important finite contribution for the UR avatars of the equations of motion. This procedure leads to the leading terms, in generic Lie algebras, which turn out to be the only 3 gluon vertex term as in \eqref{special}. It is now understood that the case of Electrodynamics obviously stands out as there is no interaction term to start with.

Next, a meticulous scrutiny of the equations of the sectors we left out, $i.e.$ \eqref{special} and \eqref{special2}, is necessary. We rejected this set, with leading Electric leg as \eqref{special} which is devoid of kinetic terms. However let us consider a particular sector of $p_0 =\mathfrak{D}-1$. In this sector the `Magnetic indices' $I,J$ only are restricted to take value 1 and the electric ones $\alpha, \beta$ run from 2 to $\mathfrak{D}$. As it happens, we can always choose a basis for the Lie algebra such that the structure constants are completely antisymmetric. In the present context this means that the left hand side of \eqref{special} vanishes identically (as $I=1=J$). Its implication in light of the discussion above is that while making an expansion in $\epsilon$ as in \eqref{expansion}, the leading non vanishing term $f^{-k_1}$($k_1 =1$ in this case.) vanishes via the conspiracy of structure constants in the particular sector of $p_0=\mathfrak{D}-1$. The sub-leading one ie $f^{-k_1+1}$ accordingly comes into play a non-trivial role here and that makes the new equation of motion:
\bea{subleadin_eqn}
 \p^{i}(\p_{i}A^{\alpha}_{0}-\p_{0}A^{\alpha}_{i}&+&gf^{\alpha}_{\hspace{.2cm}1\beta}A^{1}_{i}A^{\beta}_{0})-gf^{\alpha}_{\hspace{.2cm}\beta 1}A^{i\beta}\p_{0}A^{1}_{i}  \non\\
&& +gf^{\alpha}_{\hspace{.2cm}1\beta}A^{i1}(\p_{i}A^{\beta}_{0}-\p_{0}A^{\beta}_{i}+gf^{\beta}_{\hspace{.2cm}1\rho} A^{1}_{i}A^{\rho}_{0})=0 
\eea
replacing \eqref{special}. Here we have explicitly displayed the Lie algebra index 1, as their is only one magnetic leg for $I=1=J$.
Taking a stock of the situation we enlist the completely Magnetic \eqref{remnant1}, \eqref{remnant2} and the only mixed one corresponding to $p_0=\mathfrak{D}-1$, ie, \eqref{special2} (with $I=1$), \eqref{subleadin_eqn} as the interesting sectors. 

Also noteworthy is the status of interaction in these sectors. Interaction in the Magnetic sector appears via a term proportional to $g$ which in usual perturbation theory appears as the momentum dependent 3-gluon vertex. But more importantly the 4-gluon vertex $\sim g^2$ does appear in the mixed sector in \eqref{subleadin_eqn}. {Most importantly, this feature makes the Carrollian theory richer in structure than Galilean theories \cite{Bagchi:2014ysa, Bagchi:2015qcw}, where only $\sim g$ vertex would contribute as interaction.}


\subsubsection{Invariance of the EOM under Conformal Carroll Algebra}

One of the major goals of this work is to see whether YM theory in the UR limit respects the Carrollian conformal symmetry. Checking for invariance follows the same steps as described for the case of Electrodynamics above. One has more terms now and the additional subtlety that $\alpha, \beta, \dots$ indexed objects behave `electrically'  and $I,J ,\dots$ indexed ones transform `magnetically' under CCA. One can now simplify the analysis by using the wisdom we earned from the Electrodynamics part earlier. Invariance of those equations provided us with the fixed CCA representation, ie values of the parameters $a$ and $b$ for a particular field. Despite the fact that we had specialized our representation of CCA to 3 spatial dimensions, the entire of representation construction to finding UR projection of equations of motion is dimension independent. However, we again encounter that the spatial dimension being equal to 3 is important for this analysis. This can be seen from a sample check of vector conformal invariance of the scalar equation \eqref{remnant1} (with the special values of $a$ and $b$ being used):
\bea{}
\delta_{K_l} \left( \p^i \p_t A_i^I + g f^{I}{}_{JK} A^{iJ} \p_{t}A_i^{K}\right) =-(D-4)[\p_{t}A^{I}_{l}
-gf^{I}_{~JK}A^{J}_{i}\p_{t}A^{K}_{i}]
\eea
Similar to the relativistic case, the UR limits also respect the restriction to 3 spatial dimensions for conformal invariance. This is necessary both for Electrodynamics and YM theory. Invariance of the equations of motion \eqref{remnant1}, \eqref{remnant2} under the other generators of CCA including the infinite number of super-translations can be checked in the manner described above and the outcome is in the affirmative with the appropriate values of $a,b$. So, we again observe that the interesting sets of equations describing YM in Carrollian setting are invariant under the infinite generators of CCA.

\subsection{The lessons of this section}
In this section we have looked at the construction of the UR limit of electrodynamics and Yang Mills theories. UR electrodynamics contains two distinct sectors, $viz.$ the electric and the magnetic; and the formulation of the theory is very similar to its Galilean cousin discussed in \cite{Bagchi:2014ysa}. When we look at the generalisation to Yang-Mills theory, along lines of the discussion of the Galilean Yang-Mills theory \cite{Bagchi:2015qcw}, we expect the emergence of several different sectors depending on the individual scaling of each pair of gauge fields $\{ A^a_t, A^a_i \}$. But in contrast to the Galilean case, we found that except for the magnetic sector and a particular mixed sector, the equations of motion in the other sectors did not have kinetic terms and hence we discarded them. We included the mixed sector in our analysis as the leading equations of motion identically vanished and the sub-leading one had proper kinetic terms. This feature can be more explicitly understood in the UR version of $SU(2)$ YM theories and we have devoted an entire appendix (Appendix A) to this theory.  We investigated the symmetries of the equations of motion for these Carrollian gauge theories and found that there is the emergence of infinite dimensional Carrollian conformal symmetry in $D=4$ in all the sectors that we discussed. 
In conclusion, we wish to stress that the discussed examples would be the prototypes of theories that are holographically dual to 5d flat-space and live on its null boundary.

\section{Conclusions}

\subsection*{Summary of results}

In this paper, we have studied aspects of field theories which are putative examples of theories holographically dual to asymptotically Minkowski spacetimes. We started out with a review of some of the successes of the construction of holography for flat spacetimes in the 3d bulk -- 2d boundary case and then moved on to the more physically relevant 4d flat spacetimes. We made remarks about the highest weight representations of the BMS$_4$ algebra and, considering 3d field theories with this symmetry structure, saw that if we constructed the two and three point function of primary operators in the theory, the answers matched with that of relativistic 2d CFTs. This boiled down to the fact that the states were labelled with two Virasoro generators $L_0, \bar{L}_0$ and the supertranslations did not label the highest weight states indicating that the highest weight representations of the BMS$_4$ algebra (here we assume that the only possible non-zero central terms are the ones in the Virasoro sub-algebra, in keeping with Jacobi identities) reduce to the modules of the two copies of the Virasoro algebra. 
This hinted at a statement that 4d Minkowski spacetimes may be dual to a 2d relativistic CFT. The more conservative statement is that the 3d field theory putative dual to 4d Minkowski spacetime has 2 and 3-point correlation functions which are identical to that of a 2d relativistic CFT. 

We then moved on to general dimensions and how the ultra-relativistic or Carrollian limit on standard relativistic CFTs can be understood and how one can give these contracted algebras an infinite dimensional lift. We reviewed the procedure in field theoretic dimensions 2 and 3 and proposed a way to uplift an abelian sub-algebra of the finite contracted algebra in $D=4$, thereby generating the infinitely extended supertranslations. We then discussed the highest weight representation of this general infinite dimensional Conformal Carollian Algebra. We pointed out differences of the CCA representations with the GCA, the most stark of which was the fact that the states could only be labelled by the dilatation and the rotation generators as opposed to the GCA where we had a choice between scale-spin and scale-boost primary states on which the GCA modules were built. 

\bigskip

The second part of the paper dealt with examples of field theories which realise the CCA as their symmetry algebra. Here we first constructed the Carrollian limit of Electrodynamics and then generalised our analysis to Carrollian Yang-Mills theory. We saw that in keeping with the analysis in the similar non-relativistic limit of the theories, there are different Carrollian limits that one can take on these theories which lead to different sectors of the parent theory. In $D=4$, we found that conformal Carrollian structures arose as symmetries of the equations of motion of all these theories. The infinite dimensional supertranslations which we had proposed earlier turned out to be symmetries of these equations of motion as well. 

For general UR Yang-Mills theories, we considered only the ``vanilla" magnetic sector and a particular mixed sector. These were the only ones which   
had proper kinetic terms. The appearance of the mixed sector was subtle and required the vanishing of the leading order equations of motion due to antisymmetry of the structure constants. A more explicit version of this is worked out in Appendix A, where we construct the details of the UR limit of the $SU(2)$ Yang-Mills theory. If one did not discard \refb{special} for the lack of a kinetic term, the analysis of the symmetries would again reveal invariance under the extended infinite dimensional CCA. It may be possible to treat \refb{special} as some form of a constraint and hence keep these different sectors in our analysis. This requires further investigation which we will postpone for later work. 

Recently, the analysis of symmetries of non-relativistic Electrodynamics has been revisited in \cite{Festuccia:2016caf}. Here they find that the symmetry that arises in the non-relativistic regime of Maxwell's theory is even richer than the infinite dimensional GCA. Their symmetry algebra contains the GCA as a sub-algebra but in addition also contains an infinite dimensional $U(1)$ current algebra. This additional symmetry originates in the fact that there can be separate scalings for the space $D_{space} = x_i \p_i $ and time directions $D_{time} = t \p_t $. Following similar logic, it may be interesting to explore if the infinite dimensional BMS symmetry of the Carrollian Electrodynamics and Yang-Mills that we have discovered in this paper is further enhanced and if so, whether there is any relation to the other extensions of the BMS group.  

\subsection*{Future directions}

There are numerous immediate directions that we wish to explore following this work. Let us briefly mention some of them. We would like to add matter to our Carrollian systems and hence it is important to explore the scalar and Dirac equations in more detail and understand how to take the UR limit on theories like scalar electrodynamics and electrodynamics with fermionic fields. 

Building an action principle for these Carrollian theories along the lines of \cite{Bergshoeff:2015sic} is something we wish to pursue. We would also like to use the dual of the Newton-Cartan formulation outline in e.g. \cite{Duval:2014uoa} to revisit our present analysis in order to present a more geometric picture of the theories we have discussed. 

One of the most interesting amongst the future directions is the generalisation of the present analysis to $\mathcal{N}=4$ Supersymmetric Yang-Mills (SYM) theory. $\mathcal{N}=4$ SYM is conformally invariant even quantum mechanically in $D=4$, unlike electrodynamics and Yang-Mills theory. So it is expected that this quantum conformal invariance will be present in the UR limit and indeed would be enhanced to an infinite dimensional conformal Carrollian invariance. Similar to our motivations in the Galilean case, the appearance of infinite dimensional symmetries in the UR limit may be indicative of integrability. If so, this will be a new integrable sector in $\mathcal{N}=4$ SYM, different from the planar limit. In the context of the Holographic principle, we would like to link our analysis and its extension to $\mathcal{N}=4$ SYM to the flat limit of the best known example of holography,- $viz.$ Maldacena's AdS$_5$/CFT$_4$ correspondence. All of the above is work in progress. 

The conformal Carrollian symmetry arises in the study of the tensionless limit of string theory as a residual gauge symmetry on the string worldsheet. This is true for $D=2$ where the isomorphism with the 2d GCA and known structures there have been useful in uncovering some novel aspects of the theory previously unexplored \cite{Bagchi:2013bga, Bagchi:2015nca, Bagchi:2016yyf}. When one looks at the theory of tensionless membranes, the higher dimensional Carrollian groups discussed in this work may be similarly arise on the world-volume of these objects.  

It is possible to extend the analysis of Carrollian symmetries to study the ultra-relativistic regime of non-linear theories like Born-Infeld theories and their non-abelian generalisations. Appropriately modified versions of these theories have been speculated to be of use in tachyon cosmology \cite{Gibbons:2003gb}. It would be of interest to explore these theories in more detail. 

\bigskip

\bigskip

\section*{Acknowledgements}
We would at first list to thank Anandita De for collaboration on the initial part of this paper. It is a pleasure to also acknowledge discussions with Shankhadeep Chakrabortty, Daniel Grumiller, Jelle Hartong, Hong Liu and Joan Simon. We also thank various funding agencies for supporting this research: the Fulbright foundation and the Max Planck Society (AB), Department of Science and Technology Govt of India (RB), University Grants Commission and the NAMASTE fellowship (AM). We also warmly acknowledge the hospitality of various universities and institutes during the course of this work: AEI Potsdam, University of Groningen, Vienna University of Technology, the Simon Center for Geometry and Physics at Stony Brook, ULB Brussels. 

\newpage

\appendix
\section*{Appendix}
\section{Ultra Relativistic Yang-Mills: SU(2) case}
In this section, we will be interested in looking at the ultra relativistic limit of $SU(2)$ Yang-Mills theory. Here, we have the existence of skewed limits, over and above the Electric and Magnetic limits in the ultra relativistic electrodynamics. The reason is the presence of three different gauge fields in this case.  This leads to four distinct limits instead of two in the case of the U(1) theory. 
\subsection{EEE: Electric sector}
We begin by looking at the electric limit, where all the gauge fields transform in the same way.\\
\\
\textbf{Scaling:} All the gauge fields transform as
\be{ureeom} A_{i}^{a}\rightarrow \epsilon A_{i}^{a},\hspace{.4cm}
A_{t}^{a}\rightarrow A_{t}^{a}\ee\\
\textbf{Equations of motion:}
The equations of motion for this limit can be found by considering the scaling of spacetime coordinates along with (\ref{ureeom}):
\be{ureleom} \partial^{i}\partial_{i}A^{a}_{t}
-\partial^{i}\partial_{t}A^{a}_{i}=0,\\
 ~~\partial_{t}\partial_{t} A^{a}_{j}- \partial_{t}\partial_{j} A^{a}_{t}=0\ee
\\
\textbf{Gauge invariance:}
Our next step will  be to check the gauge invariance of the equations of motion.
Along with the scaling of $A^{a}$'s, we also have to scale the $\alpha^{a}$ as 
\be{} \alpha^{a}\rightarrow \epsilon\alpha^{a}\ee
After taking the limit, we get
\be{urgel} A_i^{a}  \rightarrow A_i^{a}+ \frac{1}{g}\partial_i \alpha^{a},\hspace{.4 cm}  A_t^{a} \rightarrow A_t^{a}+ \frac{1}{g}\partial_t \alpha^{a}\ee
The equations of motion comes out to be invariant under these transformations.  
\\
\\
\textbf{Conformal Carrollian symmetry of EOM:} 
Before looking into the conformal carrollian symmetry of the equations of motion, we should briefly state the scale-spin representations of CCA. \\
The scale-spin representations of CCA is determined by the set
$\lbrace \Delta, \Sigma, a, b\rbrace$. For each set of scalar and vector primaries $\lbrace A^{a}_{t},A^{a}_{i}\rbrace$, we assign a specific vector valued quantities $(a^{a}, b^{a})$.
For the present case, we have
\be{} \lbrace (a_1 ,b_{1}),(a_2,b_2),(a_3,b_3) \rbrace = \lbrace (0,1),(0,1),(0,1) \rbrace   \ee
 We can now show the invariance of equations of motion under the action of dilatation and special conformal transformation. Under $D$,
\be{} \partial^{i}\partial_{i}[D,A_{t}^{a}]-\partial^{i}\partial_{t}[D,A_{i}^{a}]=0,~~
\partial_{t}\partial_{t}[D,A_{i}^{a}]-\partial_{t}\partial_{i}[D,A_{t}^{a}]=0
\ee 
The equations of motion comes out to be invariant under scale transformation.\\
 Checking the invariance under special conformal transformation gives:
\bea{} (\partial.\partial)[K,A_{0}^{a}] -\partial_{i}\partial_{t}[K,A_{i}^{a}]=0\\
(\partial.\partial)[K_{i},A_{0}^{a}]-\partial_{k}\partial_{t}[K_{i},A_{k}^{a}]=(D-4)\partial_{t}A_{i}^{a}\\
\partial_{t}\partial_{t}[K,A_{i}^{a}]-\partial_{t}\partial_{i}[K,A_{0}^{a}]=0\\
\partial_{t}\partial_{t}[K_{i},A_{k}^{a}]-\partial_{t}\partial_{k}[K_{i},A_{0}^{a}]= -(D-4)\delta_{ki}\partial_{t}A_{0}^{a}\eea
We see that the EOM are invariant under $K$ in all dimensions, but only invariant under
$K_{i}$ in $D = 4$.
\\
\\
\textbf{Infinite Carrollian conformal symmetry of EOM:} Unlike relativistic counterparts, the equations (\ref{ureleom}) exhibit an infinite dimensional symmetry. To find the infinite dimensional symmetry, we will follow the same analysis used as above. We first check the transformations under $M^{m_1,m_2,m_3}$:
\bea{} \partial^{i}\partial_{i}[M^{m_1,m_2,m_3},A^{a}_{t}]
-\partial^{i}\partial_{t}[M^{m_1,m_2,m_3},A^{a}_{i}]=0,\\
 ~~\partial_{t}\partial_{t}[M^{m_1,m_2,m_3}, A^{a}_{j}]- \partial_{t}\partial_{j}[M^{m_1,m_2,m_3}, A^{a}_{t}]=0\eea   
So we see that the Electric EOM of UR $SU(2)$ Yang-Mills theory has an infinite dimensional
symmetry under $M^{m_1,m_2,m_3}$ and this is true in all dimensions.

\subsection{EEM: Skewed sector 1}  
In the first skewed limit,  $A^{1,2}$ scales electrically whereas  $A^{3}$ scales magnetically. 
\\
\\
\textbf{Scaling:} The gauge fields transform as:
\be{urIIIs} A_{t}^{1,2}\rightarrow  A_{t}^{1,2}, \hspace{.2cm} A_{i}^{1,2}\rightarrow \epsilon A_{i}^{1,2},~
A_{t}^{3}\rightarrow \epsilon A_{t}^{3}, \hspace{.2cm} A_{i}^{3}\rightarrow A_{i}^{3}\ee
\\
\textbf{Equations of Motion:}
For this limit, the equations of motion are given as
\bes\label{eomeem}
\bea{}
\p^{i}(\p_{i}A^{2}_{t}-\p_{t}A^{2}_{i}+gA^{3}_{i}
A^{1}_{t})+ gA^{i3}(\partial_{i} A_{t}^{1}-\partial_{t} A_{i}^{1}-gA^{3}_{i}A_{t}^{2}) +gA^{i1}\p_{t}A^{3}_{i}=0\label{eemeom1}
\\
\p^{i}(\p_{i}A^{1}_{t}-\p_{t}A^{1}_{i}-gA^{3}_{i}
A^{2}_{t})- gA^{i3}(\partial_{i} A_{t}^{2}-\partial_{t} A_{i}^{2}+gA^{3}_{i}A_{t}^{1}) -gA^{i2}\p_{t}A^{3}_{i}=0\label{eemeom2}\\
\p_{t}(\p_{t}A^{2}_{j}-\p_{j}A^{2}_{t}-gA^{1}_{t}A^{3}_{j})-gA^{1}_t \p_{t}A^{3}_{j}=0\label{eemeom3}\\
\p_{t}(\p_{t}A^{1}_{j}-\p_{j}A^{1}_{t}+gA^{2}_{t}A^{3}_{j})
+gA^{2}_t \p_{t}A^{3}_{j}=0\label{eemeom4}\\
\p^{i}\p_{t}A^{3}_{i}=0,\hspace{.2cm} \p_{t}\p_{t}A^{3}_{j}=0\label{eemeom5}\eea
\ees
\textbf{Gauge Invariance:} 
For the gauge transformation of the fields we shall use (\ref{urIIIs}) along with the scaling of $\alpha^{a}(a=1,2,3)$  as: 
\be{urIIIgi} \alpha^{3}\rightarrow \epsilon^{2} \alpha^{3}, \hspace{.2cm}
 \alpha^{1,2}\rightarrow \epsilon \alpha^{1,2}\ee
Gauge invariance in this limit reads
\bea{} A^{1}_{t}\rightarrow A^{1}_{t}+\frac{1}{g}\partial_{t}\alpha^{1},\hspace{.4 cm}
 A^{1}_{i}\rightarrow A^{1}_{i}+\frac{1}{g}\p_{i}\alpha^{1}-A^{3}_{i}\alpha^{2}\\
 A^{2}_{t}\rightarrow A^{2}_{t}+\frac{1}{g}\partial_{t}\alpha^{2},\hspace{.4cm}
A^{2}_{i}\rightarrow A^{2}_{i}+\frac{1}{g}\p_{i}\alpha^{2}+A^{3}_{i}\alpha^{1}\\
A^{3}_{t}\rightarrow A^{3}_{t}+\frac{1}{g}\p_{t}\alpha^{3}+A^{1}_{t}\alpha^{2}-A^{2}_{t}\alpha^{1},\hspace{.4cm}
 A^{3}_{i}\rightarrow A^{3}_{i}\eea
The equations of motion remains invariant under the given gauge transformations.
\\
\\
\textbf{Conformal Carrollian symmetry of EOM:} To find the symmetries
of the equations of motion, we need is the set of
vectors (a, b) which fix the details of the representation theory. The values are given as
\be{eemc} \lbrace (a_1,b_1),(a_2,b_2),(a_3,b_3) \rbrace = \lbrace (0,1),(0,1),(1,0)\rbrace \ee
Checking the invariance of equations of motion under scale transformation :
\bea{}
\p_{t}\p_{t}[D,A^{2}_{j}]-\p_{t}\p_{j}[D,A^{2}_{t}]-2g[D,A^{1}_{t}\p_{t}A^{3}_{j}]-
g[D,A^{3}_{j}\p_{t}A^{1}_{t}]=\non\\
-\frac{1}{2}(D-4)[2g A^{1}_{t}\p_{t}A^{3}_{j}+ gA^{3}_{j}\p_{t}A^{1}_{t}]\\
\p_{t}\p_{t}[D,A^{1}_{j}]-\p_{t}\p_{j}[D,A^{1}_{t}]+2g[D,A^{2}_{t}\p_{t}A^{3}_{j}]+ g[D,A^{3}_{j}\p_{t}A^{2}_{t}]=\non\\
\frac{1}{2}(D-4)[2g A^{2}_{t}\p_{t}A^{3}_{j}+gA^{3}_{j}\p_{t}A^{2}_{t}]\\
\p^{i}\p_{t}[D,A^{3}_{i}]=0,\hspace{.2cm} \p_{t}\p_{t}[D,A^{3}_{j}]=0\eea
Similarily, for (\ref{eemeom2}) and (\ref{eemeom1}), under $D$, we have:
\be{} \delta_{D}(\ref{eemeom2})= -\frac{1}{2}(D-4)[2gA^{3}_{i}\p_{i}A^{2}_{t}+gA^{2}_{t}\p_{i}A^{3}_{i}
-gA^{3}_{i}\p_{t}A^{2}_{i}
+gA^{2}_{i}\p_{t}A^{3}_{i}+2g^{2}A^{3}_{i}A^{3}_{i}A^{1}_{t}]\non\ee
\be{} \delta_{D}(\ref{eemeom1})= \frac{1}{2}(D-4)[2gA^{3}_{i}\p_{i}A^{1}_{t}+gA^{1}_{t}\p_{i}A^{3}_{i}
-gA^{3}_{i}\p_{t}
A^{1}_{i}
+gA^{1}_{i}\p_{t}A^{3}_{i}-2g^{2}A^{3}_{i}A^{3}_{i}A^{2}_{t}]\non\ee
The equations of motion comes out to be invariant under scale transformation in $D=4$.\\
Checking the invariance under $K$ and $K_{i}$ gives:
\bea{} 
\partial_{i}\partial_{t}[K,A_{i}^{3}]=0,\hspace{.2cm} \partial_{i}\partial_{t}[K_{l},A_{i}^{3}]=-(D-4)\partial_{t}A_{l}^{3}\\
\partial_{t}\partial_{t}[K,A_{j}^{3}]=0,\hspace{.2cm} \partial_{t}\partial_{t}[K_{l},A_{j}^{3}]=0\eea
The equations of motion (\ref{eomeem}) trivially invariant under $K$. But under $K_{l}$, we have
\bea{}
\p_{t}\p_{t}[K_{l},A^{2}_{j}]-\p_{t}\p_{j}[K_{l},A^{2}_{t}]-2g[K_{l},A^{1}_{t}\p_{t}A^{3}_{j}]- g[K_{l},A^{3}_{j}\p_{t}A_{t}^{1}]=\non\\
-(D-4)[\delta_{lj}\p_{t}A^{2}_{t}+2gx_{l}A^{1}_{t}\p_{t}
A^{3}_{j}+gx_{l}A^{3}_{j}\p_{t}A^{1}_{t}]\\
\p_{t}\p_{t}[K_{l},A^{1}_{j}]-\p_{t}\p_{j}[K_{l},A^{1}_{t}]+2g[K_{l},A^{2}_{t}\p_{t}A^{3}_{j}]+ g[K_{l},A^{3}_{j}\p_{t}A_{t}^{2}]=\non\\
-(D-4)[\delta_{lj}\p_{t}A^{1}_{t}-2gx_{l}A^{2}_{t}
\p_{t}A^{3}_{j}-gx_{l}A^{3}_{j}\p_{t}A^{2}_{t}]  
\eea  
Similarily, for (\ref{eemeom1}) and (\ref{eemeom2}), under $K_{l}$, we have:
\bea{} 
\delta_{K_{l}}(\ref{eemeom1})= (D-4)\p_{t}A^{2}_{l}+(D-4)x_{l}[2gA^{3}_{i}\p^{i}A^{1}_{t}-gA^{3}_{i}\p_{t}A^{1}_{i}
+gA^{1}_{t}\p^{i}A^{3}_{i}\non\\+gA^{1}_{i}\p_{t}A^{3}_{i}-2g^{2}A^{3}_{i}A^{3}_{i}
A^{2}_{t}]+(D-4)gA^{3}_{l}A^{1}_{t}
\eea
\bea{}
\delta_{K_{l}}(\ref{eemeom2})= (D-4)\p_{t}A^{1}_{l}-(D-4)x_{l}[2gA^{3}_{i}\p^{i}A^{2}_{t}-gA^{3}_{i}\p_{t}A^{2}_{i}
+gA^{2}_{t}\p^{i}A^{3}_{i}\non\\ +gA^{2}_{i}\p_{t}A^{3}_{i}+2g^{2}A^{3}_{i}A^{3}_{i}
A^{1}_{t}]-(D-4)gA^{3}_{l}A^{2}_{t}
\eea
The equations of motion are only invariant in $D=4$ under $K_{l}$.
\\
\\
\textbf{Infinite Carrollian conformal symmetry of EOM:} To check the infinite dimensional symmetry of the equations of motion, we have to use the transformation of the fields under  $M^{m_1,m_2,m_3}$ and (\ref{eemc}). Invariance under $M^{m_1,m_2,m_3}$ is given by:
\be{} \delta_{M}(\ref{eemeom1})=0,~\delta_{M}(\ref{eemeom2})=0,~\delta_{M}(\ref{eemeom3})=0,~\delta_{M}(\ref{eemeom4})=0,~ \delta_{M}(\ref{eemeom5})=0 \nonumber \ee
The equations of motion in this limit are also invariant under infinite dimensional CCA in all dimensions.

\subsection{EMM: Skewed sector 2}
In the second skewed limit, $A^{1}$ scales electrically whereas $A^{2,3}$ scales magnetically. 
\\
\\
\textbf{Scaling:} In this limit, the gauge fields scales as
\be{}\label{urIIs} A_{t}^{2,3}\rightarrow \epsilon A_{t}^{2,3},\hspace{.2cm} A_{i}^{2,3}\rightarrow A_{i}^{2,3},~
A_{t}^{1}\rightarrow A_{t}^{1}, \hspace{.2cm} A_{i}^{1}\rightarrow \epsilon A_{i}^{1}\ee\\
\textbf{Equations of motion:} 
The equations of motion for skewed sector 2 are given as
\bea{emmeom} \p^{i}\p_{t}A^{2,3}_{i}=0,\hspace{.2cm}
g[A^{i2}\p_{t}A^{3}_{i}-A^{i3}\p_{t}A^{2}_{i}]=0\\
\p_{t}\p_{t}A^{2,3}_{j}=0,\hspace{.2cm}
\p_{t}(\p_{t}A^{1}_{j}-\p_{j}A^{1}_{t})=0\eea
\\
\textbf{Gauge Invariance:} 
In order to have the gauge invariance of the equations of motion, we have to use the gauge transformation of the fields (\ref{urIIs}) along with the gauge parameter scaled as
\bea{}\label{urIIgi} \alpha^{2,3}\rightarrow \epsilon^{2} \alpha^{2,3}, \hspace{.2cm}
 \alpha^{1}\rightarrow \epsilon \alpha^{1}\eea
Gauge transformations of the gauge fields are written as
\bea{} A^{3}_{t}\rightarrow A^{3}_{t}+\frac{1}{g}\partial_{t}\alpha^{3},\hspace{.4 cm}
 A^{3}_{i}\rightarrow A^{3}_{i}\\
 A^{2}_{t}\rightarrow A^{2}_{t}+\frac{1}{g}\partial_{t}\alpha^{2},\hspace{.4cm}
A^{2}_{i}\rightarrow A^{2}_{i}\\
A^{1}_{t}\rightarrow A^{1}_{t}+\frac{1}{g}\p_{t}\alpha^{1},\hspace{.4cm}
 A^{1}_{i}\rightarrow A^{1}_{i}+ \frac{1}{g}\partial_{i}\alpha^{1}\eea
Under these transformations, the equations of motion remains invariant.
\\
\\
\textbf{Conformal Carrollian symmetry of EOM:} 
To obtain the symmetries of the equations of motion, we would require the set of vectors $\lbrace a,b\rbrace$ that fix the details of the representation theory. The values are given by
\be{} \lbrace (a_1,b_1),(a_2,b_2),(a_3,b_3)\rbrace =\lbrace (0,1),(1,0),(1,0) \rbrace \ee
Under scale transformation, the invariance of equations of motion is written as
\bea{} \p^{i}\p_{t}[D,A^{2,3}_{i}]=0,~~
\p_{t}\p_{t}[D,A^{2,3}_{j}]=0,\\
\p_{t}\p_{t}[D,A^{1}_{j}]-\p_{t}\p_{j}[D,A^{1}_{t}]=0\eea
The equations of motion comes out to be invariant under scale transformation in all dimensions.
Checking the invariance under $K$ and $K_{l}$: 
\bea{} 
\partial_{i}\partial_{t}[K,A_{i}^{2,3}]=0,\hspace{.2cm} \partial_{i}\partial_{t}[K_{l},A_{i}^{2,3}]=-(D-4)\partial_{t}A_{l}^{2,3}\\
\partial_{t}\partial_{t}[K,A_{j}^{2,3}]=0,\hspace{.2cm} \partial_{t}\partial_{t}[K_{l},A_{j}^{2,3}]=0\\
\partial_{t}\partial_{t}[K,A_{i}^{1}]-\partial_{t}\partial_{j}[K,A^{1}_{t}]=0\\
\partial_{t}\partial_{t}[K_{i},A_{j}^{1}]-\partial_{t}\partial_{j}[K_{i},A^{1}_{t}]= -(D-4)\delta_{ji}\partial_{t}A_{0}^{1}
\eea
In this limit, the equations of motion are invariant under $K_{i}$ only in $D=4$.\\
\\
\textbf{Infinite Carrollian conformal symmetry of EOM:} We would follow the same analysis as above to find the invariance of equations of motion under infinite CCA. Under $M^{m_1,m_2,m_3}$,
\bea{} \p^{i}\p_{t}[M^{m_1,m_2,m_3},A^{2,3}_{i}]=0,~
[M^{m_1,m_2,m_3},gA^{i2}\p_{t}A^{3}_{i}-gA^{i3}\p_{t}A^{2}_{i}]=0\\
\p_{t}\p_{t}[M^{m_1,m_2,m_3},A^{2,3}_{j}]=0,\hspace{.2cm}
\p_t \p_t[M^{m_1,m_2,m_3}, A^{1}_{j}]-\p_t \p_{j}[M^{m_1,m_2,m_3},A^{1}_{t})]=0\eea
The equations of motion are invariant in all dimensions.

\subsection{MMM: Magnetic sector}
In magnetic limit, all gauge fields $( A^{1},A^{2},A^{3} )$ scales in the same manner.
\\
\\
\textbf{Scaling:} The gauge fields transforms as 
\be{}\label{urmll} A_{i}^{a}\rightarrow  A_{i}^{a},\hspace{.4cm}
A_{t}^{a}\rightarrow \epsilon A_{0}^{a}\ee\\
\textbf{Equations of motion:} 
The equations of motion in the magnetic limit are
\be{} \partial^{i}\partial_{t}A^{a}_{i}
+g\varepsilon^{abc}A^{ib}\partial_{t}A^{c}_{i}=0,~~
\partial_{t}\partial_{t} A^{a}_{j}=0\ee\\
\textbf{Gauge Invariance:} The gauge transformation of the fields can be obtained if we take into account the scaling of potentials (\ref{urmll}) along with the scaling of $\alpha^{a}$: 
\be{} \alpha^{a}\rightarrow \epsilon^{2} \alpha^{a}\ee
In this limit, the gauge transformation becomes
\be{} A_i^{a}  \rightarrow A_i^{a},\hspace{.4 cm}
A_t^{a} \rightarrow A_t^{a}+ \frac{1}{g}\partial_t \alpha^{a}\ee
The equations of motion comes out to be invariant under the gauge transformations.
\\
\\
\textbf{Conformal Carrollian symmetry of EOM:} 
To get the symmetries of the equations of motion, we have to consider the set of vectors $\lbrace a,b \rbrace$ that put together the details of the representation theory. The values of vectors are  
\be{} \lbrace(a_1,b_1),(a_2,b_2),(a_3,b_3)\rbrace =\lbrace(1,0),(1,0),(1,0) \rbrace \ee
Under scale transformation, the invariance is given as
\be{} \partial^{i}\partial_{t}[D,A_{i}^{a}]+g\varepsilon^{abc} [D,A^{b}_{i}\p_{t}A^{c}_{i}]=\frac{1}{2}(D-4)g\varepsilon^{abc}A^{b}_{i}\p_{t}A^{c}_{i},~~
\partial_{t}\partial_{t}[D,A_{i}^{a}]=0
\ee
The equations of motion comes out to be invariant under scale transformation in $D=4$.
Under $K$ and the $K_{l}$, we have
\bea{} \partial^{i}\partial_{t}[K_{l},A_{i}^{a}]+g\varepsilon^{abc}[K_{l},A^{b}_{i}\partial_{t}A^{c}_{i}]=-(D-4)[\partial_{t}A_{l}-g\varepsilon^{abc}x_{l}A^{b}_{i}\partial_{t}A^{c}_{i}]
\\
\partial^{i}\partial_{t}[K,A_{i}^{a}]+g\varepsilon^{abc}[K,A^{b}_{i}\partial_{t}A^{c}_{i}]=0 ,~
\partial_{t}\partial_{t}[K,A_{j}^{a}]=0,\hspace{.2cm} \partial_{t}\partial_{t}[K_{l},A_{j}^{a}]=0
\eea
The equations of motion are invariant under $K$ in all dimensions, whereas they are invariant under $K_{l}$ in $D=4$.
\\
\\
\textbf{Infinite Carrollian conformal symmetry of EOM:} Checking the invariance under Infinite CCA gives,
\be{} \partial^{i}\partial_{t}[M^{m_1,m_2,m_3},A^{a}_{i}]
+g\varepsilon^{abc}[M^{m_1,m_2,m_3},A^{ib}\partial_{t}A^{c}_{i}]=0,~~
\partial_{t}\partial_{t}[M^{m_1,m_2,m_3}, A^{a}_{j}]=0\ee
The equations of motion are invariant in all dimensions.

\end{document}